\documentclass{article}
\usepackage[english]{babel}
\usepackage[margin=1in]{geometry}
\usepackage{multicol}
\usepackage[math]{blindtext}

\newcommand{\vecn}[1]{\hat{#1}}
\usepackage{graphicx}
\usepackage{epsfig}
\usepackage{amsmath}
\usepackage{empheq} 
\usepackage{amsfonts}
\usepackage{amssymb}
\usepackage{subfig}

\begin{document}

{\centering

{\bfseries\Large Qualitative Analysis of the Time-Frequency Signature Induced by a Reflected L-Band Signal from Time Evolving Sea 
Surfaces\bigskip}

Arnaud~Coatanhay and Alexandre~Baussard\\
  {\itshape
\textsuperscript{1}ENSTA Bretagne / Lab-STICC (UMR CNRS 6285), Brest, France\\
\normalfont (Dated: December 17, 2013)
  }
}

\begin{abstract}
Passive remote sensing techniques have become more and more popular for
detection and characterization purposes. The advantage of using the Global
Navigation Satellite Systems (GNSS) are the well known signals emitted and the
availability in most areas on Earth. In the present paper, L-Band signals
(including GNSS signals) are considered for oceanographic purposes. The main
interest in this contribution is the analysis of the signal reflected by an
evolving sea surface using time-frequency transforms. The features which occur
in this domain are examined in relation to the physical phenomena:
interaction of the electromagnetic waves with the moving sea surface.
\end{abstract}

\section{Introduction}

For a decade, the passive remote sensing techniques based upon
electromagnetic sources of opportunity have become more and more popular for
detection and characterization purposes. The most significant reason for this
enthusiasm lies in the fact that the passive systems take advantage of the
numerous existing electromagnetic emitting sources to come up with new
approaches for remote sensing or surveillance applications. As a matter of
fact, among all the possible sources of opportunity, the Global Navigation
Satellite Systems GNSS (GPS, GALILEO, GLONASS,...) appear among the most
relevant solutions since the electromagnetic signals emitted (in L-Band) are
reliable, available all over the world, deterministic and perfectly known (for direct paths).

It is worth noting that the use of GNSS signals for passive observation has been
applied to a significant extent to the field of oceanography since the
year 2000. Indeed, several studies have been carried out to
estimate oceanographic parameters using the reflected GNSS signals measured from
ground-based systems, see for example \cite{Belmonte_2006, Marchan_2008}, or
airborne systems, see for example \cite{Garrison_2011, Yang_2008, Voo_2010,
Zavorotny_2000a}. These systems can be used for the measurement of wind
velocity and wind direction \cite{Garrison_2002, Komjathy_2000, Yang_2008,
Zavorotny_2000}, surface roughness and its effects \cite{Garrison_2011,
Thompson_2002, Voo_2010} or even to estimate the salinity \cite{Marchan_2008}.

For airborne and long range systems, the precision in time involves that the resolution cell easily exceeds a few hundred
square meters in surface area. Then, the scattering by each individual sea wave can not be discriminated and described.
In these cases, the sea surface is considered as a very large stochastic rough surface
induced by the complex interaction between the fluid and the wind. In this
context, the reflected GNSS is mostly seen as a tool to extract
the statistical characteristics of a maritime environment (root mean square
deviation of the sea surface height for instance).

For ground-based systems (see \cite{Belmonte_2006} for instance), the studies mainly focus on the delay/Doppler analysis of
the GPS signals. The modulation of the GPS signal makes the sharp analysis on long coherent period quite
difficult and limits the resolution in the time frequency domain.

The recent Maritime Opportunity Passive Systems
(MOPS) \cite{Coatanhay_conf_2008} research
project considers a local maritime domain observed in the vicinity of the sea
surface. The prime objective of this project is to obtain a very high resolution analysis (very long coherent period and  
high precision in time) of the GPS phase for the direct and reflected signals in order to estimate the sea surface movements.
From preliminary studies it appears that the obtained delay/Doppler maps show quite complex patterns that remain difficult
to interpret. However, these patterns clearly suggests that this passive GNSS system should be able to detect the movement
of individual sea waves.

The general purpose of this paper is closely related to this research project. It  investigates the possibility of
observing the movement and the deformation of a sea surface from the reflected signal in L-Band. In practice, we investigate the
connections between a deterministic time evolving surface and the time-frequency representation of the signal
scattered by this surface. The idea is to take advantage of these representations to
extract oceanographic parameters from Doppler and micro-Doppler signatures. In a broad outline, a sea surface can be seen as a sum
of `sinusoidal' surfaces at several scales. Basically, it can be expected that the large scale,
corresponding to the gravitational waves, will be related to the global motion
of the sea and the small scales, corresponding to the capillary waves directly
induced by the wind, to the roughness of the sea surface. 

In this paper, we will concentrate on the interpretations of the time-frequency
signatures of the signal scattered by the evolving sea surface. These
interpretations should be of a great importance for complementary works dealing with the estimation of oceanographic
parameters. We can expected to estimate, as already noted, the movements and the deformations of a sea surface (i.e. the
local sea state and the sea roughness) but also the local wind velocity and direction.

First, our approach consists in generating time evolving (two-dimensional) surfaces and computing the field of a plane wave in
L-Band scattered by these surfaces. The numerical model is based upon the Method of Moment (MoM) which is a standard
approach used to quantify the scattering by sea surfaces, in L-Band. Then, the
time-frequency signatures of the scattered signal are obtained using
Wigner-Ville representations. Finally, making comparisons
with the signatures produced by canonical surfaces (moving sinusoids for
instance), we provide interpretations of the complex signatures corresponding
to more realistic sea surfaces.

The sequel of the paper is organized as follows. In Section \ref{sec:em_model}
the electromagnetic model used to compute the scattered electromagnetic
field is introduced. Section \ref{sec:sea_model} describes the model applied to
generate realistic evolving sea surfaces. Then, Section \ref{sec:TFoverview}
presents a short overview of the
time-frequency representations used in this paper. Sections \ref{sec:sinus} and
\ref{sec:sea} give an interpretation of the time-frequency representations of the
signal reflected from canonical and modeled sea surfaces.
Finally, Section \ref{sec:conclusion} gives some concluding remarks and proposes
some future works dealing with the extraction of the characteristics of the sea surface from the time-frequency representations.

\section{Electromagnetic model} \label{sec:em_model}
\begin{figure}[!ht] 
\centering
\includegraphics[height=9cm,angle=-90]{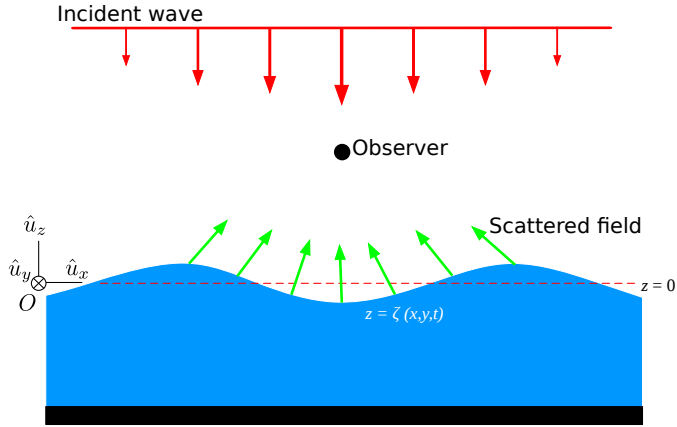}
\caption{Geometrical configuration.}
\label{fig:configuration}
\end{figure}

Roughly speaking, the GNSS signal can be considered as a plane Right
Hand Circular Polarization (RHCP) incident wave. For the sake of simplicity,
we assume in the following that the satellites are at the Zenith position, so
the incident wave propagates in the Nadir direction, see Figure
\ref{fig:configuration}.

In this configuration, the electromagnetic modeling
consists in the evaluation of the scattering of a plane RHCP wave by a
dielectric interface. As a matter of fact, due to the salinity of the sea water,
the surface is assumed to be a Perfect Electric Conductor (PEC).
It is noteworthy that a circular polarized plane wave can be split into two
linear polarized plane waves: Transverse Magnetic (TM) and Transverse Electric
(TE) polarization. An appropriate approach to modeling the scattering by TM and TE
polarized wave is to consider the Boundary Integral Equation Method (BIEM) and
the estimation using functional basis: Method of Moments (MoM).

For the TM polarization ($\vec{E}=E\vecn{u}_y$), the modeling of the
scattering is based upon the Electric Field Integral Equation (EFIE)
\cite{Gibson}:
\begin{equation}\label{e:EFIE}
E_{y}^{inc}(\vec{\rho}\,)=i\omega_{em}\mu\int_{\Gamma}J_{TM}(\vec{\rho}\,
')G\left(\vec{\rho},\vec{\rho}\,'\right)d\vec{\rho}\,',
\end{equation}
where $E_{y}^{inc}$ is the incident field, $\mu$ is the permeability of free
space, $\Gamma$ represents the sea surface, $\omega_{em}$ is the angular
frequency of the electromagnetic wave, $J_{TM}$ is the surface
current, $G(\vec{\rho},\vec{\rho}\,')$ is the 2D Green function,
$\vec{\rho}\,'$ and $\vec{\rho}$ are the source and observation points
on the sea surface ($\vec{\rho}\,', \vec{\rho}\in\Gamma$).
It is worth noting that
the incident wave is known and is assumed to be tapered to avoid numerical
truncation effects. So, the amplitude $E_y^{inc}$ of incident field will be
given by \cite{Thorsos_art_1988}:
\begin{equation}
E_y^{inc}\left(x\right)=e^{
-ik z\left(1+v_t\left(x
\right)\right) } e^{-\frac{x^2}{g_t^2}},
\end{equation}
where $g_t$ is the tapering parameter and the additional factor in the phase
$v_t$
is:
\begin{equation}
 v_t\left(x\right)=\frac{\left[2\frac{x^2}{
g_t^2 }
-1\right]}{\left(kg_t\right)^2}.
\end{equation}

The 2D Green function $G(\vec{\rho},\vec{\rho}\,')$ is also perfectly known, and
the only unknown function of the integral equation (\ref{e:EFIE}) is the surface
current $J_{TM}$. When the surface is computed, the scattered field is
determined using the following relation:
\begin{equation}\label{e:MFIE_bis}
E_{y}^{scat}(\vec{\rho}_{obs}\,)=-i\omega_{em}\mu\int_{\Gamma}J_{TM}(\vec{\rho}
\,')G\left(\vec{\rho}_{obs},\vec{\rho}\,'\right)d\vec{\rho}\,',
\end{equation}
where $\vec{\rho}_{obs}$ is the position of the observer above the sea surface.

For the TE polarization, the electrical field $\vec{E}$ belongs to the plane
($O,\vecn{u}_x,\vecn{u}_z$) and the magnetic field is in the form ($\vec{H}=H_y
\vecn{u}_y$). So the modeling of the scattering in TE polarization is based upon
the Magnetic Field Integral Equation (MFIE)\cite{Gibson}:
\begin{equation}\label{e:MFIE}
H_{y}^{inc}(\vec{\rho}) =
-\frac{J_{TE}(\vec{\rho})}{2}+\int_{\Gamma}J_{TE}(\vec{\rho}\,')
\left[\hat{n}(\vec{\rho}\,')\nabla'G\left(\vec{\rho},\vec{\rho
}\,'\right)\right]d\vec{\rho}\,',
\end{equation}
where $\hat{n}\left(\vec{\rho}\right)$ is the normal vector to the sea
surface at the position $\vec{\rho}$, $\nabla'G\left(\vec{\rho}\right)$ is the
gradient related to the second variable of the Green function, $H_{y}^{inc}$ is
the incident field and $J_{TE}$ is the surface current.

In the same way as the TM polarization case, the integral equation
(\ref{e:MFIE}) determines the surface current $J_{TE}$. So, the magnetic
field $\vec{H}$ received by the observer is computed using a relation similar to
equation (\ref{e:MFIE_bis}), and then the electric field received by the
observer can be evaluated. 

Considering both TM and TE polarizations and a standard Method of Moment (MoM) approach to solve the integral equations
(\ref{e:EFIE} and \ref{e:MFIE}) (a detailed presentation can be found
in \cite{Tsang_book_2001b}), we can simulate the scattering of a GNSS signal
(RHCP plane wave) by a sea surface. 

The last point for the electromagnetic modeling that needs to be said is that
the receiver antenna of the observer is Left Hand Circular Polarized (LHCP) to
optimize the recording of the reflected signal. So the actual received signal
will be the LHCP component of the scattered field.

\section{Sea surface model} \label{sec:sea_model}

\begin{figure}[!hbt]
\begin{minipage}[b]{.99\linewidth}
  \centering
  \centerline{\epsfig{file=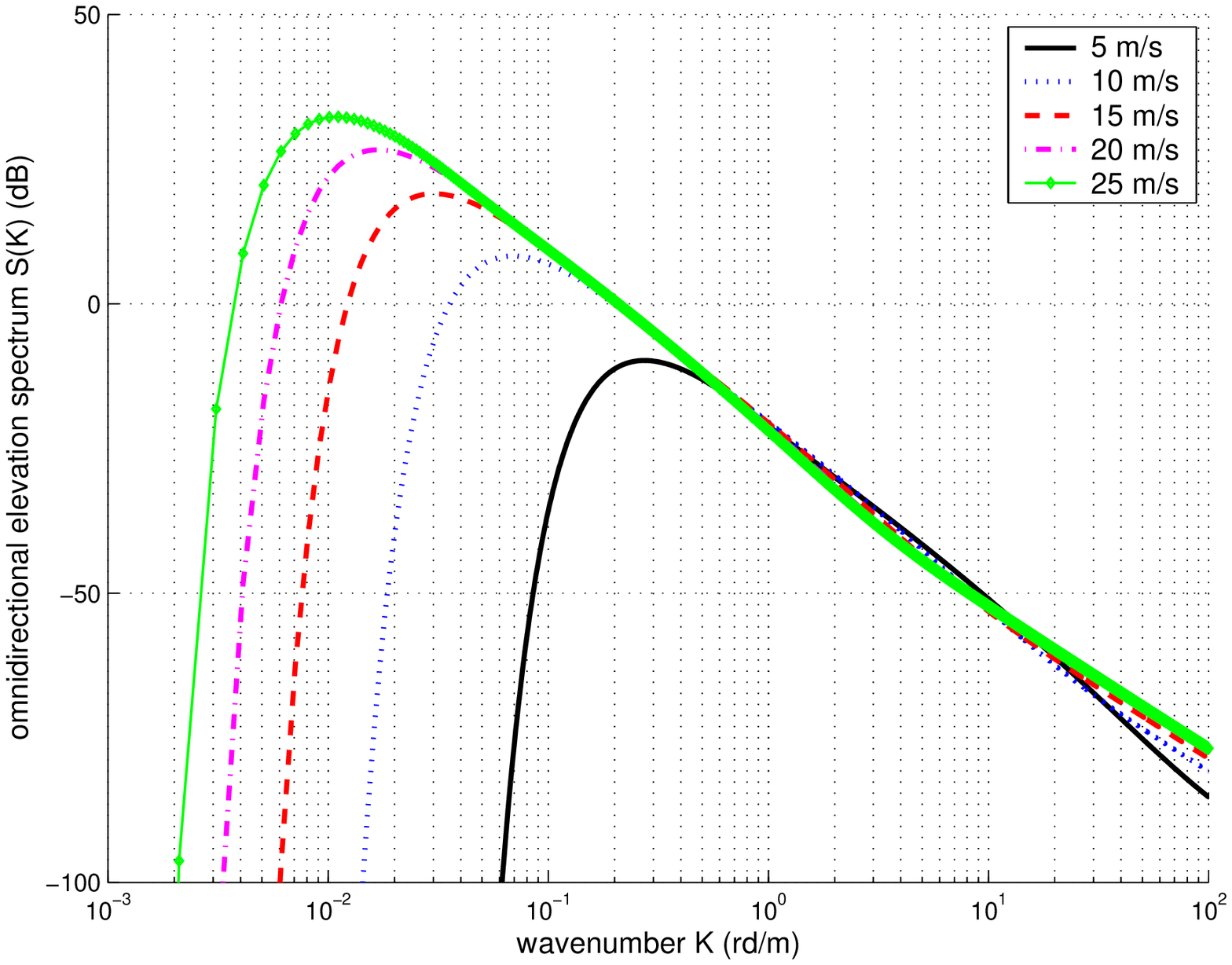,width=7cm}}
  \vspace{0.05cm}
  \centerline{(a)}
\end{minipage}
\begin{minipage}[b]{.99\linewidth}
  \centering
  \centerline{\epsfig{file=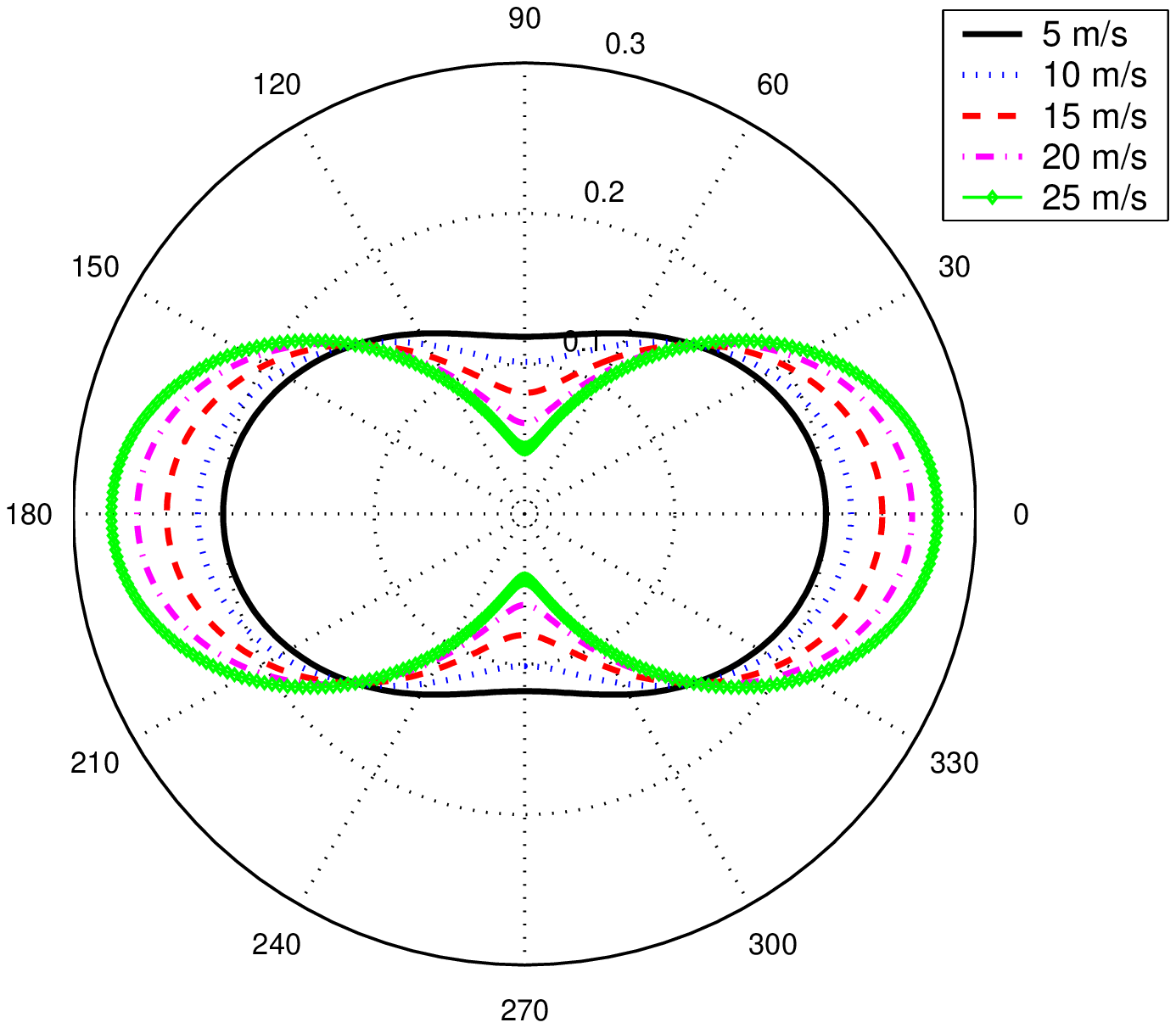,width=7cm}}
  \vspace{0.05cm}
  \centerline{(b)}
\end{minipage}
\caption{Elfouhaily sea surface spectra with different wind speeds: (a)
Omnidirectional elevation spectrum component (b) Angular function component.} 
\label{fig:elf}
\end{figure}

Since the previously described electromagnetic model estimates the
scattered field for any surface, the last step to obtain a complete
simulation is the generation of the sea surfaces. Physically, a sea surface is
induced by a complex and stochastic interaction between the fluid and the wind.
Fortunately, a statistical description of the energy spectral decomposition of
the sea surface can be approximated by empirical and/or semi-theoretical models
for different wind speeds and wind directions. One of the most consistent with
the experimental data has been developed by
Elfouhaily et al. \cite{Elfouhaily_art_1997}. This sea spectrum is in the form:
\begin{equation}\label{equa10}
    S(K_{sea},\,\phi)=M(K_{sea})f(K_{sea},\,\phi_{sea}),
\end{equation}
where $M(K_{sea})$ represents the isotropic part of the spectrum modulated by
the angular function $f(K_{sea},\,\phi_{sea})$. $K_{sea}$ and $\phi_{sea}$
respectively denote the spatial wave number and the wind direction, see Figure \ref{fig:elf}.

To generate a realistic sea surface associated to a given weather
condition (wind speed and wind direction), we expand the surface into a
continuous sum of sinusoidal curves. For each spatial wave number $K_{sea}$, the
mean amplitude is estimated using the square root of the sea spectrum and the phase
is randomly set between 0 and $2\pi$.
In practice, the convolution of the square root of the spectrum with a unitary
white Gaussian random signal generates a one-dimensional profile (a statistical
realization of the sea surface) that represents an ocean surface for given
weather conditions.

To describe the time evolution of the sea surface thus obtained, we associate a
time wave number $\omega_{sea}\left(K_{sea}\right)$ for each $K_{sea}$ using the
phase dispersion
relation:
\begin{equation}\label{e:dispersion}
   \omega_{sea}=\sqrt{g_{g}K_{sea}\left(1+\left(K_{sea}A\right)^2\right)},
\end{equation}
where $g_{g}$ is the acceleration due to gravitation and $A$ is the
amplitude of the sinusoidal curve.

Finally, we can generate a realistic time-evolving surface and using the
electromagnetic model we can also estimate the signal recorded by the receiver.
Now, we will demonstrate that these simulated signals analyzed with time-frequency
tools lead to many physical interpretations.

\section{Time-Frequency overview}\label{sec:TFoverview}

The numerical simulation of the electromagnetic waves intercepted by an
observer above a time-varying sea surface provides complex non-stationary
signals. A relevant approach to fully describe the nature of
these non-stationary signals consists in a Time-Frequency (TF) analysis.

Indeed, a Time-Frequency Distribution (TFD) shows how 
the spectral content of the signal evolves with time, thus providing an ideal
tool to dissect, analyze and interpret non-stationary signals. Most of time,
the Time-Frequency Distribution (TFD) maps the energy of a one-dimensional
time-domain signal into a two-dimensional function of time and frequency. 

In this article, we will stress the fact that the feature extraction based on
time-frequency analysis of an electromagnetic field above the sea can be
exploited to point out useful information about the signal fluctuation induced
by the movement of the surface.

A great variety of methods for obtaining a TFD have been defined, most
notably the short time Fourier transform, the wavelet transform and the
Wigner-Ville distribution. For more details the reader can refer for example to
\cite{Boashash_2003} and the Time-Frequency
Toolbox (TFTB\footnote{http://tftb.nongnu.org/}). This article deals with the Wigner-Ville distribution.

\subsection{The Wigner-Ville Distribution}
The Wigner-Ville Distribution (WVD) of a signal $y(t)$, denoted by $W_z(t,f)$, is defined as
\begin{equation}
W_z(t,f)=\int_{-\infty}^{\infty} z(t+\tau/2)z^*(t-\tau/2)e^{-j2\pi f \tau} {\rm d}\tau,
\end{equation}
where $z(t)$ is the analytic associate of $y(t)$ (i.e the Hilbert transform of the real signal $y(t)$).

The main properties of the WVD are : it is always real-valued, it preserves
time and frequency shifts and it satisfies the marginal properties.

Unfortunately, the WVD shows interference terms. These interference terms are
troublesome since they may overlap with signal terms and thus make it difficult
to visually interpret the WVD image. Basically, the interference between two
points in the time-frequency plane correspond to the appearance of a third point
located at the geometrical midpoint. Besides, the interference terms oscillate
perpendicularly to the line joining the two interfering points, with a frequency
proportional to the distance between these two points.

In order to reduce the interference terms, the Pseudo-Wigner-Ville Distribution
(PWVD) can be used. This transform is described in the following.

\subsection{The Pseudo-Wigner-Ville Distribution}
The PWVD is defined as
\begin{equation}
PW_z(t,f)=\int_{-\infty}^{\infty} h(\tau) z(t+\tau/2)z^*(t-\tau/2)e^{-j2\pi f \tau} {\rm d}\tau,
\end{equation}
where $h(\tau)$ is a regular window. 

This windowing is equivalent to a frequency smoothing of the WVD. It leads
to the attenuation of the interference terms. However, an excessive window leads to
a loss in properties and may damage the joint-time-frequency resolution.

\section{Reflected signal in the TF domain - canonical surfaces}\label{sec:sinus}
As already mentioned, the electromagnetic field scattered by a realistic
time-varying sea surface induces complex non-stationary signals received by the
observer. So, in a first phase, we suggest generating and analyzing the
received signal when the simulated sea corresponds to a canonical moving
surface (a sinusoid for instance). 

In this section, we will attempt to provides several interpretations of
the time-frequency signatures for simple or more sophisticated canonical
reflecting surfaces. This step strikes us as necessary to understand and
explain the main features obtained in the time-frequency domain for more
complex reflecting surfaces. In what follows, we will considers the increasing
complexity of the simulated reflecting surface: from the case of a simple
sinusoidal surface to the case of the linear superposition of
several sinusoidal surfaces.

\subsection{Main parameters of the canonical surfaces}
Let the sea surface be approximated, for its global shape and motion, by a simple sinusoid $s$ defined as
\begin{equation}
s(t,x)=A_{sin}\sin(-\omega_{sin} t+k_{sin} x).
\label{eq:sin}
\end{equation}
The main parameters of this surface are the amplitude $A_{sin}$, the velocity
$c_{sin}$, the angular frequency $\omega_{sin}=2\pi f_{sin}$ or the
wavelength
$\lambda_{sin}=c_{sin}/f_{sin}$ and the wavenumber $k_{sin}=2\pi /
\lambda_{sin}$. In (\ref{eq:sin}), $t$ stands for the time and $x$ for the
position. To come closer to a realistic sea model, the parameters of this
sinusoidal model should be inferred from the sea surface model
introduced in Section \ref{sec:sea_model}. 

For given weather conditions (wind speed and wind direction), the wavenumber of
the sinusoid $k_{sin}$ is fitted to the abscissa $K_{sea,max}$ of the maximum 
of omnidirectional elevation spectrum component, see Figure \ref{fig:elf}. Since
the wavenumber $k_{sin}$ is set, the angular frequency $\omega_{sin}$ can be
computed using the phase dispersion relation (\ref{e:dispersion}). The
wavelength $\lambda_{sin}$ and the velocity $c_{sin}$ can then be deduced.

To determine the amplitude of the sinusoid, we generate realistic
sea surfaces from the Elfouhaily spectrum. Then, the standard
deviation $\sigma_{sea}$ of the wave height is computed. In physical oceanography, the global characteristic is given by the
Significant Wave Height (SWH or Hs) which is defined
traditionally as the mean wave height (trough to crest) of the highest third of
the waves ($H_{1/3}$). In practice, the SWH is usually defined as four times the
standard deviation of the surface elevation: $SWH=4\sigma_{sea}$. This is
why the sinusoid amplitude is set to $A_{sin}=2\sigma_{sea}$.

The estimated parameters defining the sinusoidal surfaces are given in Table
\ref{tab:beaufort}. Note that, in this paper, only the Beaufort scales from 1 to
5 are considered.

\begin{table}
	\center
	\begin{tabular}{|c|c|c|c|c|c|}
	\hline 
	Scale &	Description & wind speed & Amplitude & Wavelength & velocity\\ 
	&	& (m/s) & (m) & (m) & (m/s) \\
	\hline
	1 & Light & 0.3-1.5 & 0.0-0.012 & 0.4-2.44 & 0.8-1.95 \\
	& air	& & & & \\ \hline			
	2 & Light & 1.6-3.3 & 0.014-0.12 & 2.77-11.8 & 2.08-4.29 \\ 
	& breeze & & & & \\ \hline
	3 & Gentle & 3.5-5.4 & 0.15-0.54 & 13.3-31.6 & 4.55-7.02 \\ 
	& breeze & & & & \\ \hline		
	4 & Moderate & 5.5-7.9 & 0.57-1.70 & 32.8-67.8 & 7.15-10.27 \\ 
	& breeze & & & & \\ \hline
	5 & Fresh & 8-10.7 & 1.77-4.23 & 69.4-124.3 & 10.4-13.9 \\
	& breeze & & & & \\ \hline	
	\end{tabular}
	\caption{Sinusoidal sea surface parameters.}
\label{tab:beaufort}
\end{table}

\begin{center}
\begin{figure}[]
  \centering
  \subfloat[]{\label{fig:edge-a}\includegraphics[width=8.5cm]{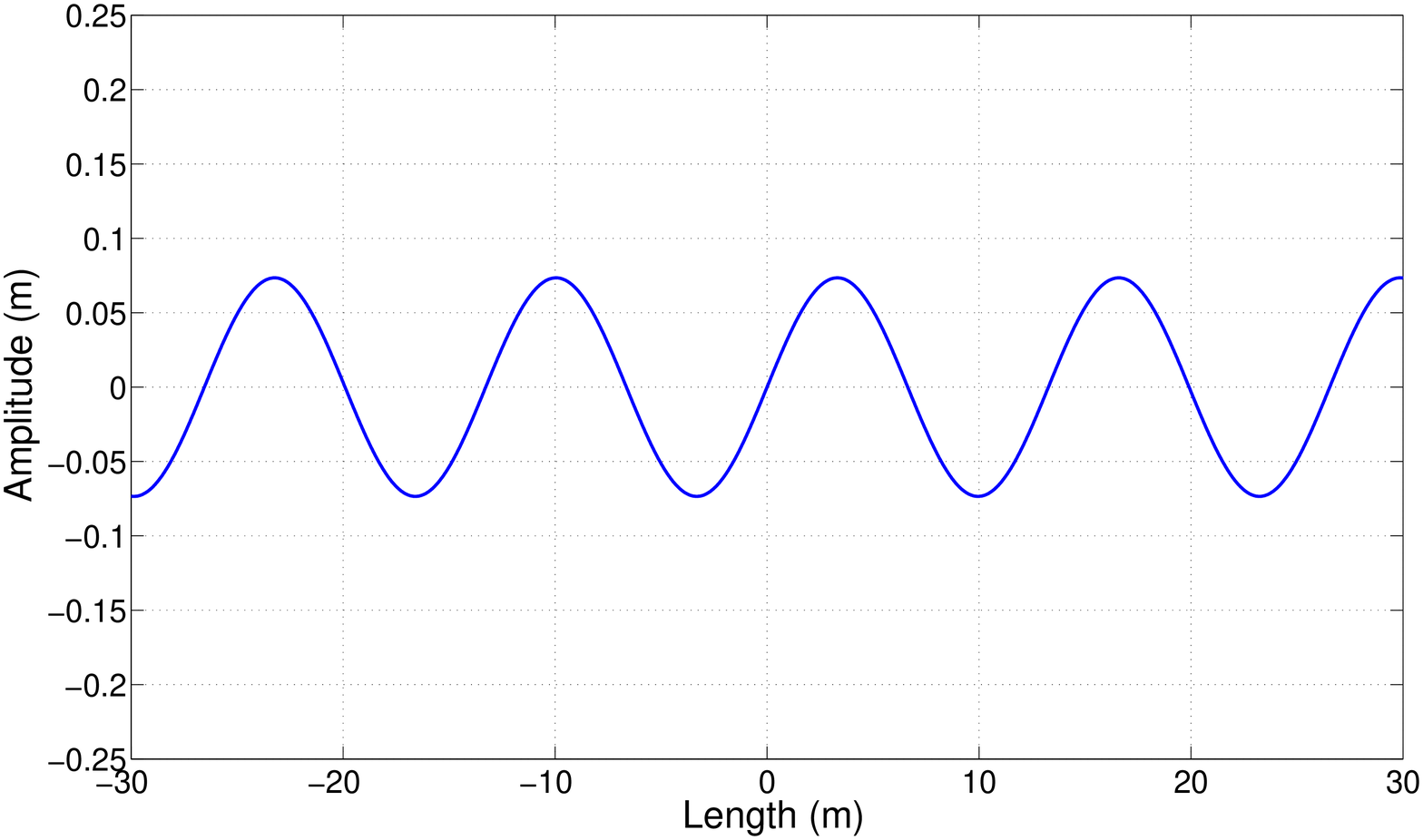}}                
  \subfloat[]{\label{fig:contour-b}\includegraphics[width=8.5cm]{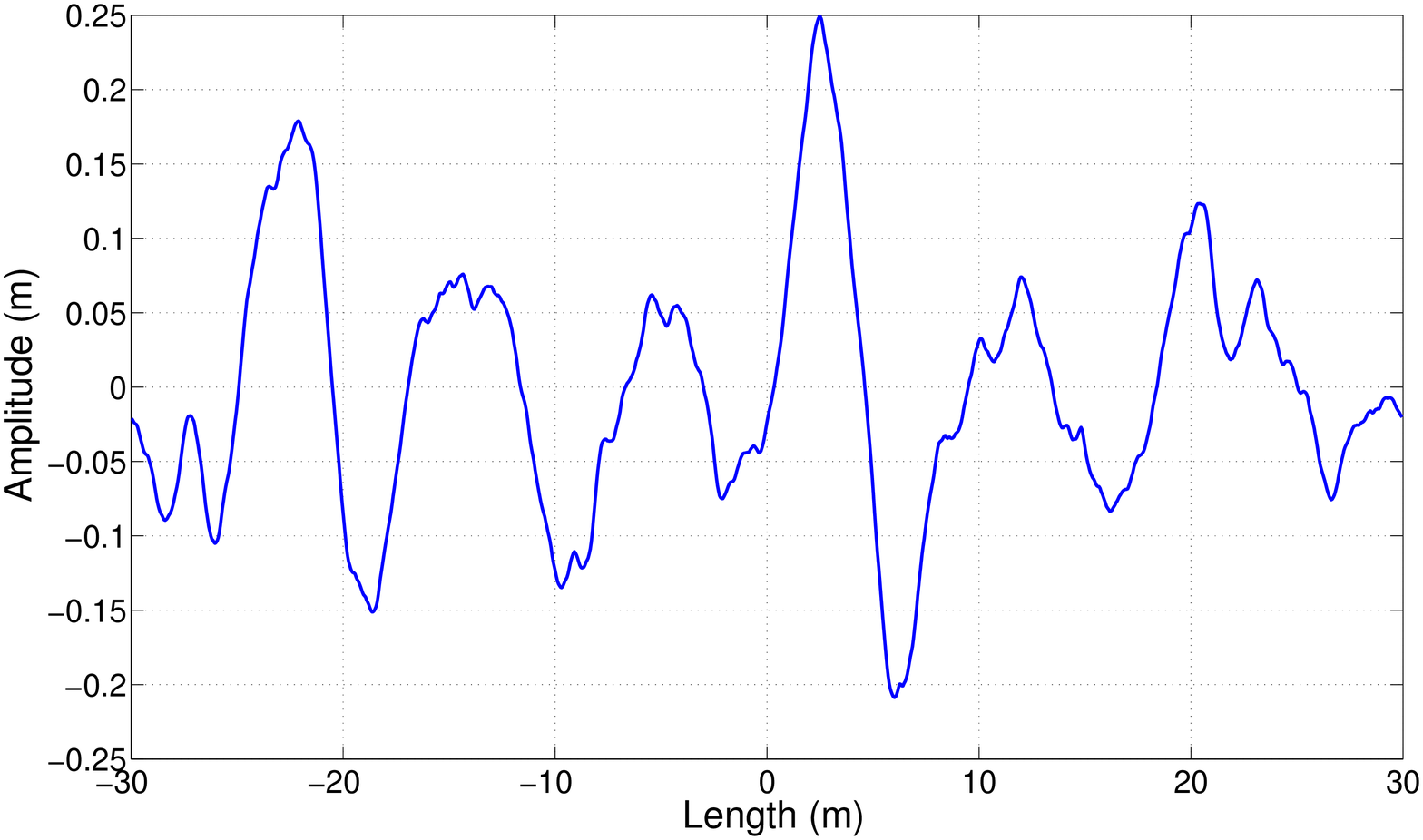}}
  \caption{\label{fig:ex_surface} Generated (a) pure sinusoidal surface and (b) sea surface for a Beaufort scale 3.}
\end{figure}
\end{center}

These simplifying assumptions and the estimated parameters have been
experimentally validated by comparing the obtained evolving sinusoidal surfaces,
for several values of the Beaufort scale, with the global shape and motion of
several sea surfaces (generated using the Section \ref{sec:sea_model} model) at the
same Beaufort scales. Figure \ref{fig:ex_surface} shows a $60$m sinusoidal
surface at time $t$ and the corresponding sea surface for scale 3. The main visual
difference is the wavelength of both surfaces. It clearly appears from Figure
\ref{fig:ex_surface} that the canonical surface corresponds to only one
wavenumber and the simulated sea surface to a mixture of a wide frequency
spectrum. However, the sinusoidal surface obtained remains broadly an
acceptable model in the first instance.

\subsection{Simulation for a pure sinusoidal surface}

\begin{figure}[!ht] 
\centering
\includegraphics[width=\textwidth,trim=1.5cm 1.5cm 1.5cm 1.5cm, clip=true]{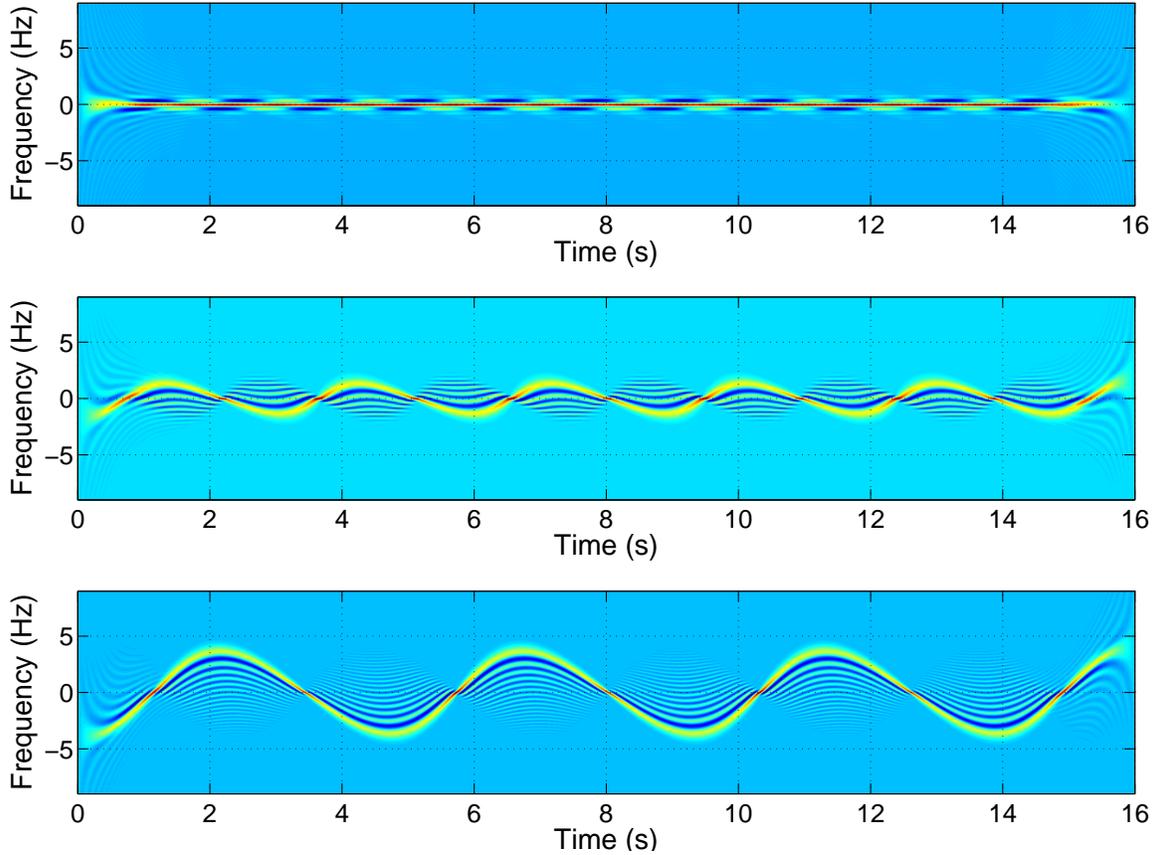}
\caption{Time-frequency representation (with same color scale) of the received signal reflected by a sinusoidal surface with
parameters corresponding to the Beaufort scales (top) 2, (middle) 3 and (bottom) 4.}
\label{fig:ex_surface_sin}
\end{figure}

Considering the previous sinusoidal surface and the
electromagnetic model presented in Section \ref{sec:em_model}, we are able
to simulate the signal measured at the receiver over a period of $16$s. Then,
applying a Pseudo-Wigner-Ville Distribution (PWVD), a two-dimensional
representation of this signal can be achieved in the time-frequency domain. Figure \ref{fig:ex_surface_sin} presents the
time-frequency representation of the received signal reflected from surfaces corresponding to the levels 2, 3 and 4 in the
Beaufort
scale.

\begin{figure}[!ht]
\centering
\includegraphics[width=\textwidth]{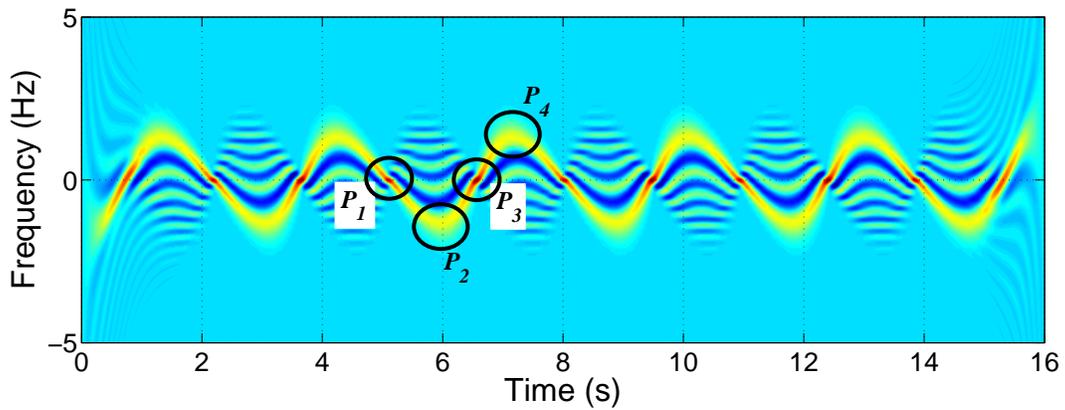}
\caption{Zoom of Figure \ref{fig:ex_surface_sin} (middle). The white circles
point out the time-frequency signature at four different times: $t_1=5.10$s,
$t_2=5.98$s, $t_3=6.56$s and $t_4=7.13$s.}
\label{fig:TF_sinus_surf_3}
\end{figure}

At first sight, the time-frequency feature seems very close to a sinusoid. This
is mainly due to the fact that the scattering phenomenon becomes time-periodic
in the case of a sinusoid surface.
Nevertheless, a closer examination shows that the time-frequency feature
looks more like a distorted sinusoid, see
Figure \ref{fig:TF_sinus_surf_3}. We must not forget that the TF
representation shows the variations of the Doppler frequency observed by the
receiver induced by the motion of the sinusoid surface. Finding the relation
between the evolving sea surface elevation and the Doppler variations is the key
challenge in analyzing the PWVD and interpreting the time-frequency
signature. 

In order to give a useful interpretation, let us denote by $P_1$, $P_2$, $P_3$
and $P_4$ four significant points observed in the PWVD of the signal received
for a sinusoid surface, see Figure \ref{fig:TF_sinus_surf_3}. The abscissa
of these points are the time delays $t_1=5.10$s,
$t_2=5.98$s, $t_3=6.56$s and $t_4=7.13$s. The points $P_2$ and $P_4$ respectively
correspond to a maximum and a minimum of the time frequency signature and the
points $P_1$ and $P_3$ correspond to two Doppler zero-crossings.

\begin{figure}[!ht]
\centering
\includegraphics[width=9cm,trim=1.5cm 1.5cm 1.5cm 1.5cm, clip=true]{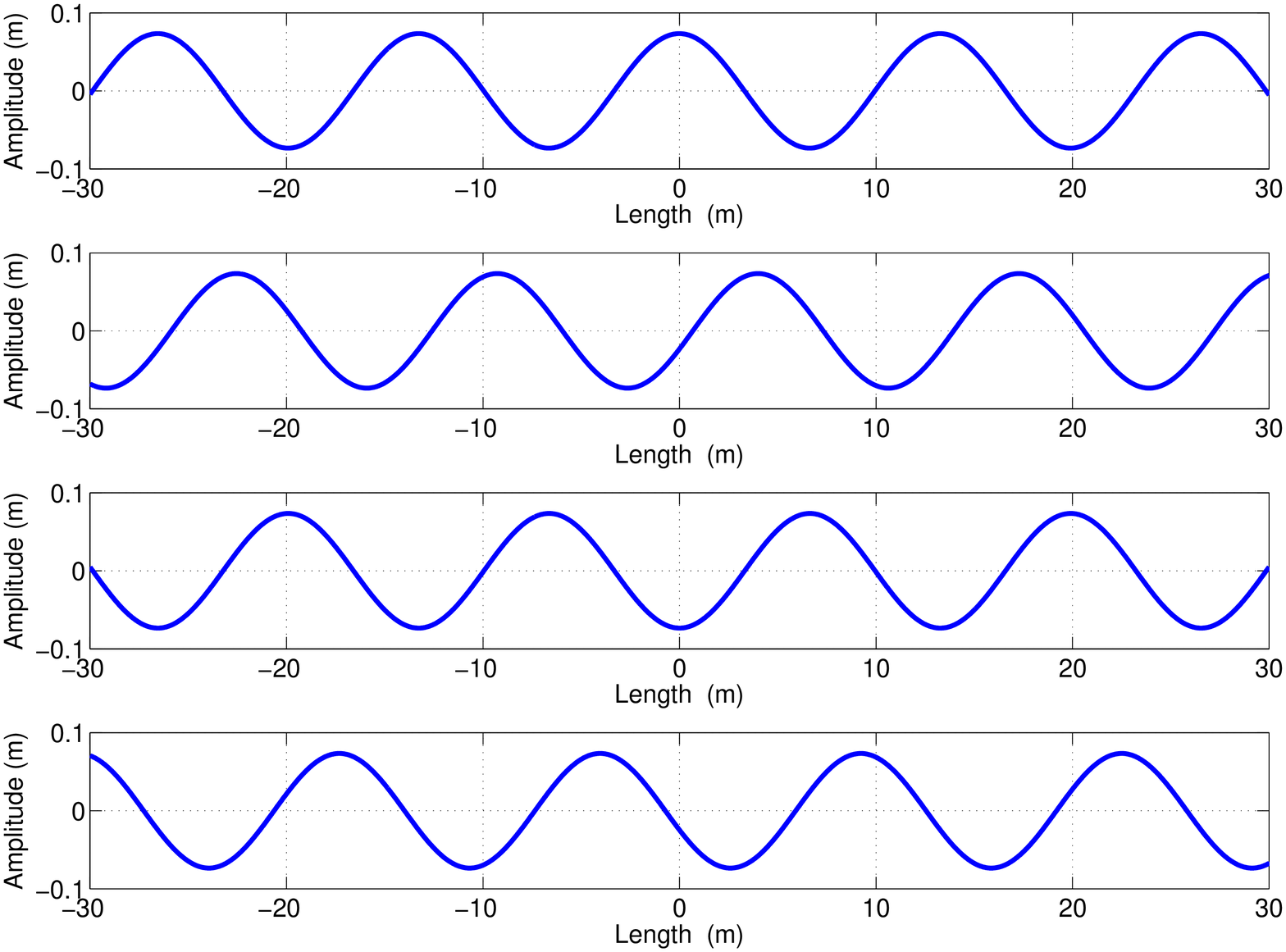}
\caption{The sinusoidal sea surfaces at the four different times (top to
bottom): $t_1=5.10$s, $t_2=5.98$s, $t_3=6.56$s and $t_4=7.13$s.}
\label{fig:sinus_surf_3}
\end{figure}

\begin{figure}[!hbt]
\begin{minipage}[b]{.99\linewidth}
  \centering
  \centerline{\epsfig{file=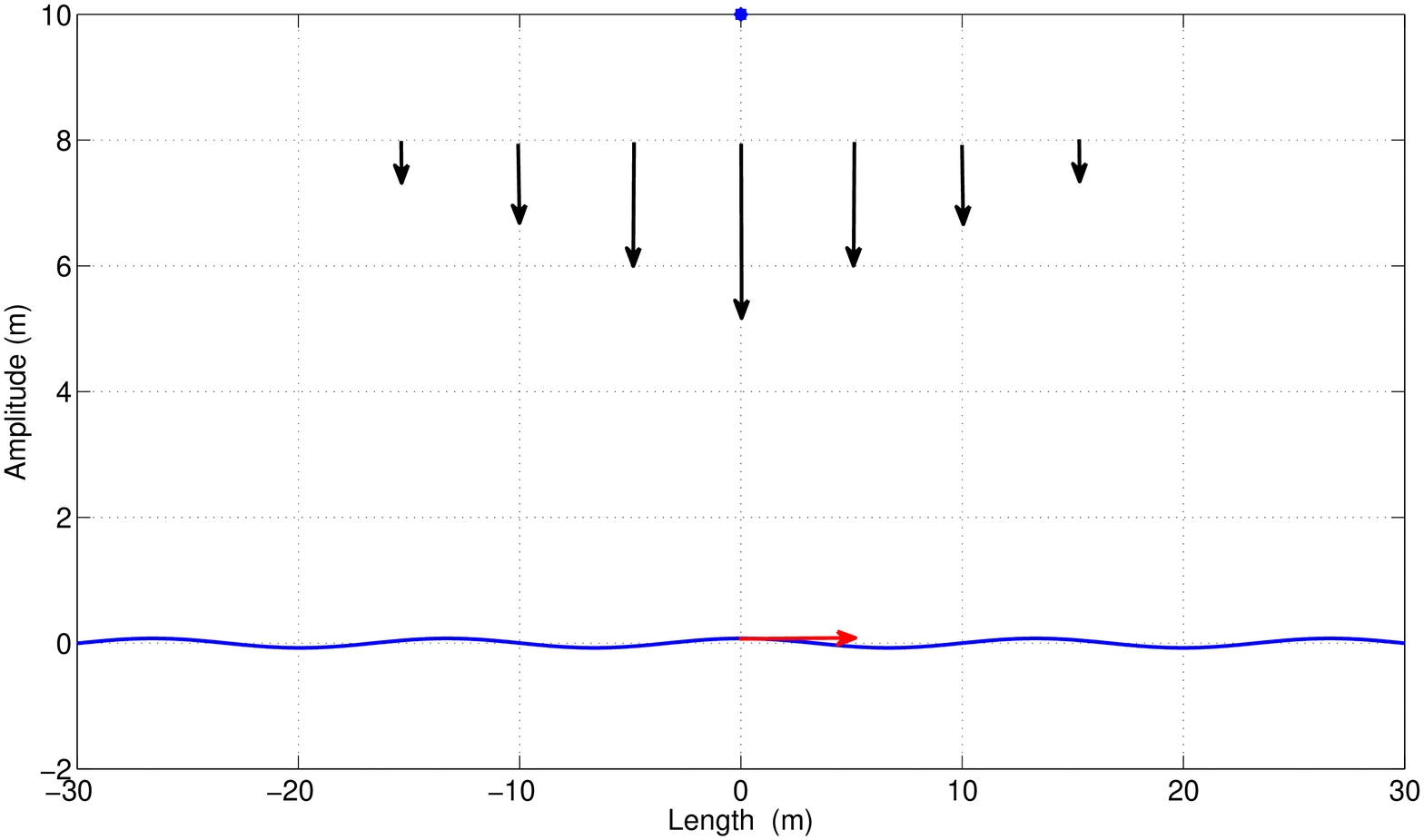,width=7cm}}
  \vspace{0.05cm}
  \centerline{(a)}
\end{minipage}
\begin{minipage}[b]{.99\linewidth}
  \centering
  \centerline{\epsfig{file=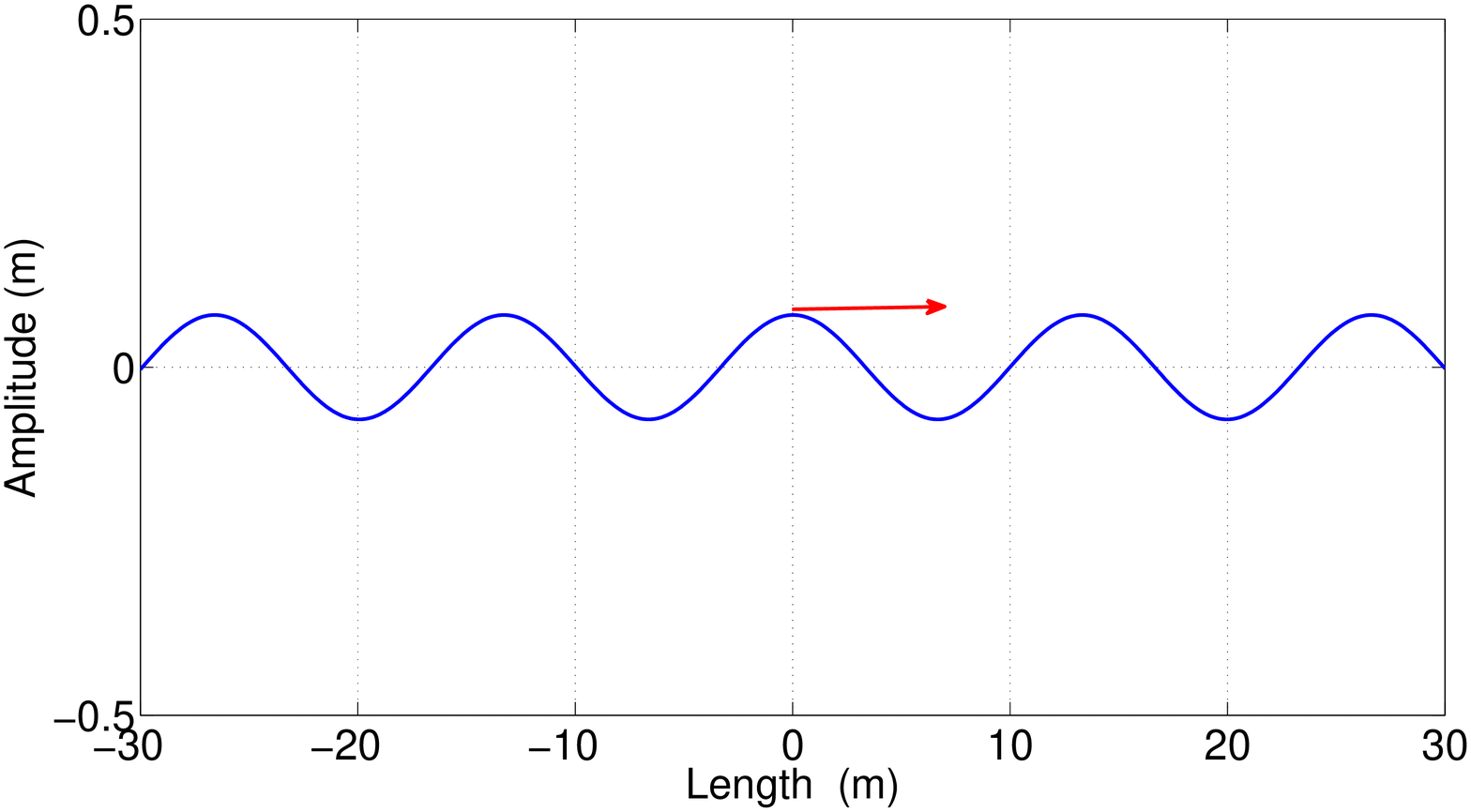,width=7cm}}
  \vspace{0.05cm}
  \centerline{(b)}
\end{minipage}
\caption{(a) Setup of the system at $t_1=5.10$s ($P_1$), (b) zoom of (a).}
\label{fig:zero_doppler}
\end{figure}

\begin{figure}[!hbt]
\begin{minipage}[b]{.99\linewidth}
  \centering
  \centerline{\epsfig{file=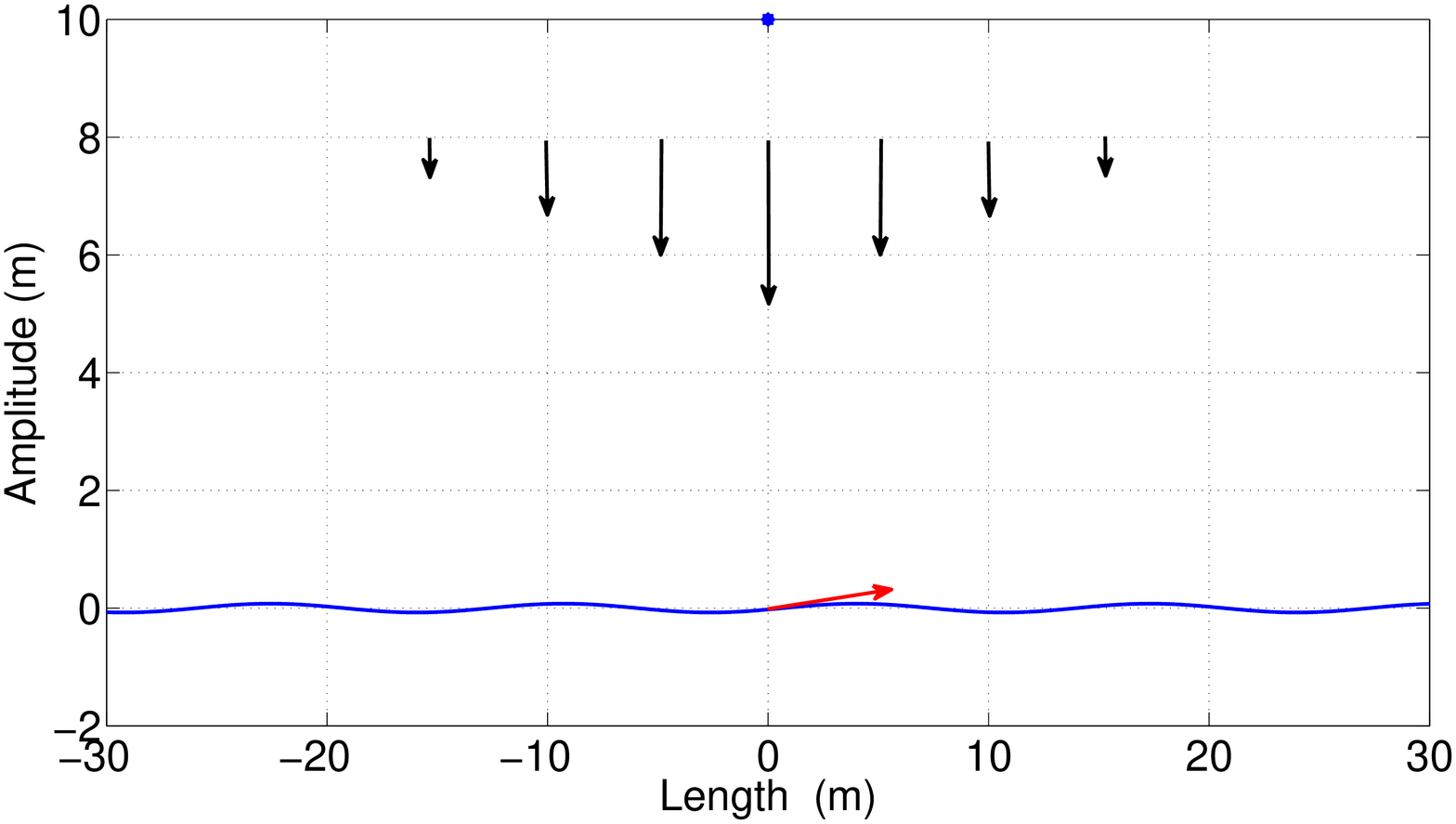,width=7cm}}
  \vspace{0.05cm}
  \centerline{(a)}
\end{minipage}
\begin{minipage}[b]{.99\linewidth}
  \centering
  \centerline{\epsfig{file=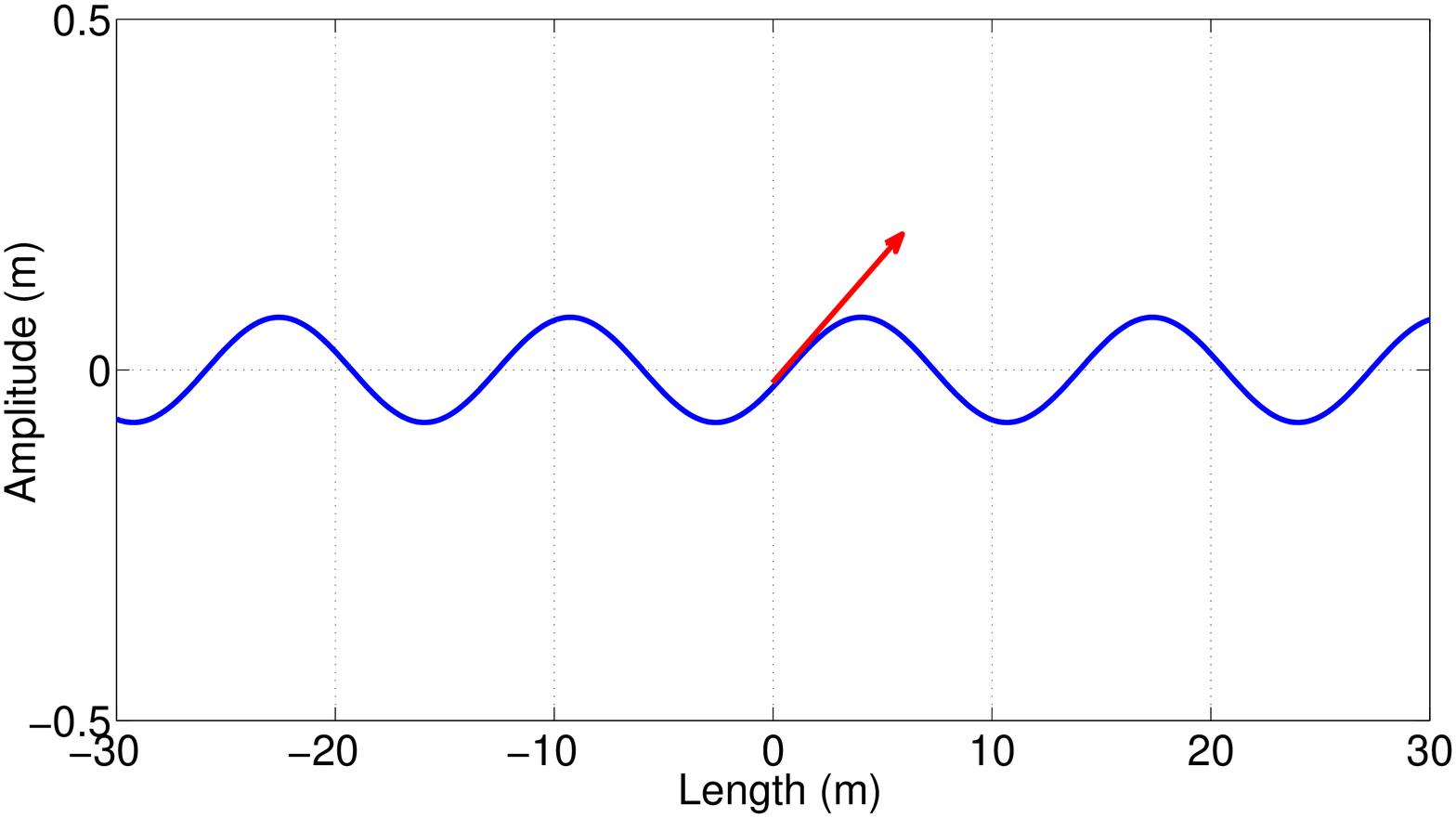,width=7cm}}
  \vspace{0.05cm}
  \centerline{(b)}
\end{minipage}
\caption{(a) Setup of the system at $t_2=5.98$s, (b) zoom of (a).}
\label{fig:max_doppler}
\end{figure}

The Doppler frequency is zero ($P_1$ and $P_3$) when a maximum or a minimum of
the sinusoidal surface is under the receiver position (see Figure \ref{fig:sinus_surf_3}). In fact, due to the tapering of the
incident field, the incident wave mainly impinges on central area of the
sea surface and using the approximation of the ray theory, the main electromagnetic contribution coming from the sea surface is
induced by the specular point situated at the top (or the
bottom) of the sea surface, see Figure \ref{fig:zero_doppler}. The velocity
associated to this specular point (red arrow in Figure \ref{fig:zero_doppler})
is strictly perpendicular to the observer direction, so the Doppler
shift is zero. Note that in Figure \ref{fig:zero_doppler}, and also in Figure
\ref{fig:max_doppler}, the black arrows stand schematically for the intensity
(length of the arrow) and direction of the incident wave.

Numerically, if we remember that the wavelength of the sinusoidal surface is $\lambda_{sea}=13.3m$, the velocity is
$c_{sea}=4.55m/s$ and the time period is $T_{sea}=2.92s$, we note that $t_3-t_1=1.46s$ exactly corresponds to half of the
$T_{sea}$ period.

On the other hand, the Doppler frequency is maximal or minimal ($P_2$ and $P_4$)
when the amplitude (see Figure \ref{fig:sinus_surf_3}) of the sinusoidal
surface above the observer is close to zero (i.e. the inflexion point of the
sinus). Figure \ref{fig:max_doppler}
illustrates this configuration and shows the velocity vector at this point.

However, the extrema of the Doppler signature do not strictly correspond
to the zero-crossings of the sea surface, and $t_4-t_2=1.15$s$\simeq 1.46$s is an
approximation of half of the $T_{sea}$ period. Thus we can
say that the Doppler signature is a distorted sinusoid.
Furthermore, a simple calculation shows that the velocity vector at the point
above the observer does not correspond to the Doppler shift at point $P_2$ or
$P_4$.

In the sequel, we show that it is possible to make a further, detailed
analysis of the time-frequency signature by drawing analogies from the ray
theory and explaining the shape of the signature by the presence of
specular points.

\begin{figure}[!ht] 
\centering
\includegraphics[width=\textwidth]{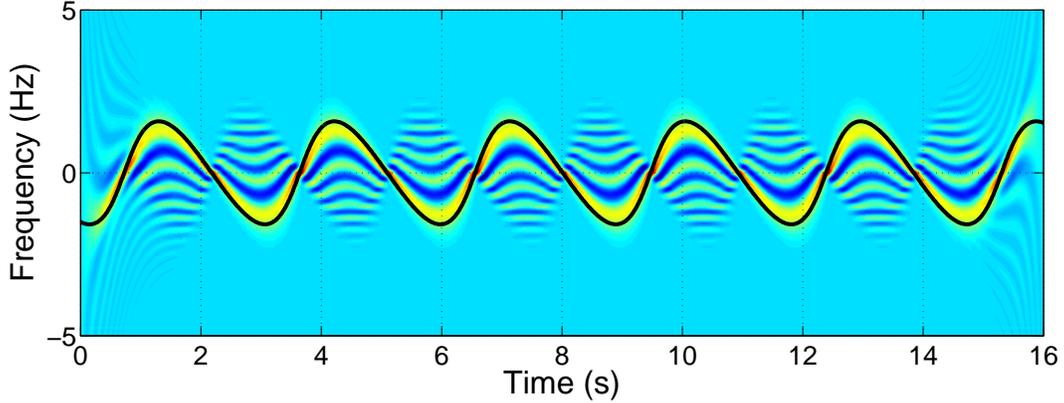}
\caption{PWVD of the received signal and (in black) Doppler frequency of the
moving specular point.}
\label{fig:sinus_spec_3}
\end{figure}

More precisely, the feature in the time-frequency domain for this kind of 
surface is due to the motion of the specular point which reflects maximum energy
in the receiver direction. To illustrate this, Figure \ref{fig:sinus_spec_3}
shows the PWVD of the received signal after reflection on the sinusoidal surface
and the evolution (black line) of the Doppler frequency due to the single
moving specular point (for the considered sinusoidal surface only one specular
point exists at each time).

In practice, the specular point is geometrically determined in the central lobe
of the incident wave and its Doppler frequency is
computed as:
\begin{equation} \label{eq:spec_fd}
f_d(t)=\frac{1}{\lambda_{em}}\frac{{\rm d}}{{\rm d} t}(R_{ET}+R_{TR}),
\end{equation}
where $R_{ET}$ and $R_{TR}$ are respectively the distances Emitter-Target and Target-Receiver.

At L-band, for a simple sinusoidal surface and the considered setup, the MoM
model and the single point source (specular point) approximation have
significant differences in quantitative terms. However, in terms of qualitative
description, both approaches provide
very similar time frequency signatures. 
This can be easily illustrated by generating the signal
$V(t)$ measured at the receiver due to the specular point considered as
a source point. This signal is obtained from
\begin{equation}
V(t)=\exp[j (2 \pi f_{em} t - k_{em} r)],
\end{equation}
where $r$ is the distance covered by the electromagnetic wave, $f_{em}$ the
frequency of the incident wave and $k_{em}$ the electromagnetic wave number.

\begin{figure}[!ht] 
\centering
\includegraphics[width=\textwidth]{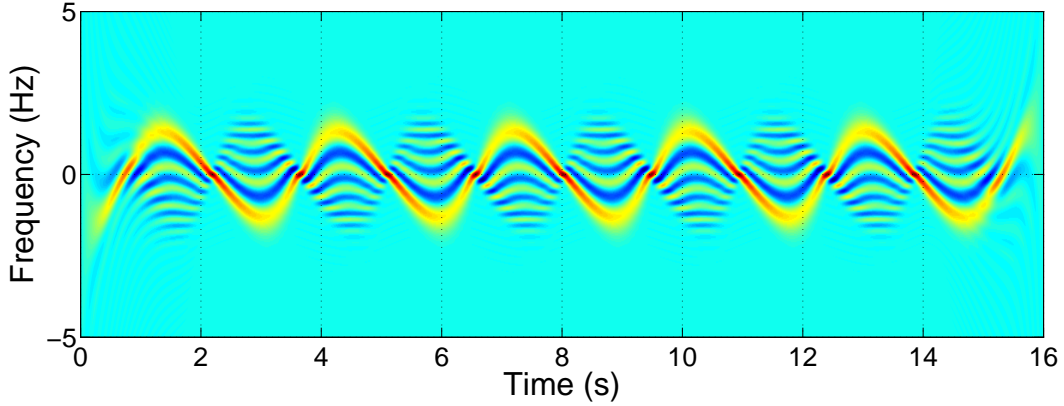}
\caption{Time-frequency representation of the signal reflected by a sinusoidal surface using the point source model. To be
compared with Figure \ref{fig:sinus_spec_3}.}
\label{fig:simu_scat}
\end{figure}

Figure \ref{fig:simu_scat} shows the PWVD obtained using this `point source
model'. The feature amplitudes are quite different since a basic source model
does not take into account the complex phenomena of attenuation.
Nevertheless, the PWVD in Figure \ref{fig:simu_scat} is very nearly comparable
to that obtained in Figure \ref{fig:sinus_spec_3}.

\begin{figure}[!ht] 
\centering
\includegraphics[width=10cm]{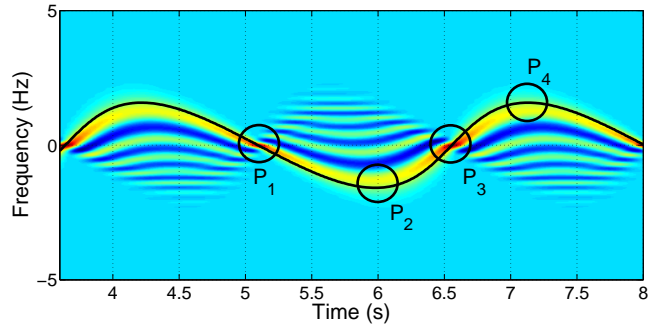}
\caption{Zoom of the PWVD of the received signal (Figure \ref{fig:sinus_spec_3}).}
\label{fig:sinus_3_zoom}
\end{figure}

\begin{figure}[htb]
\begin{minipage}[b]{.99\linewidth}
  \centering
  \centerline{\epsfig{file=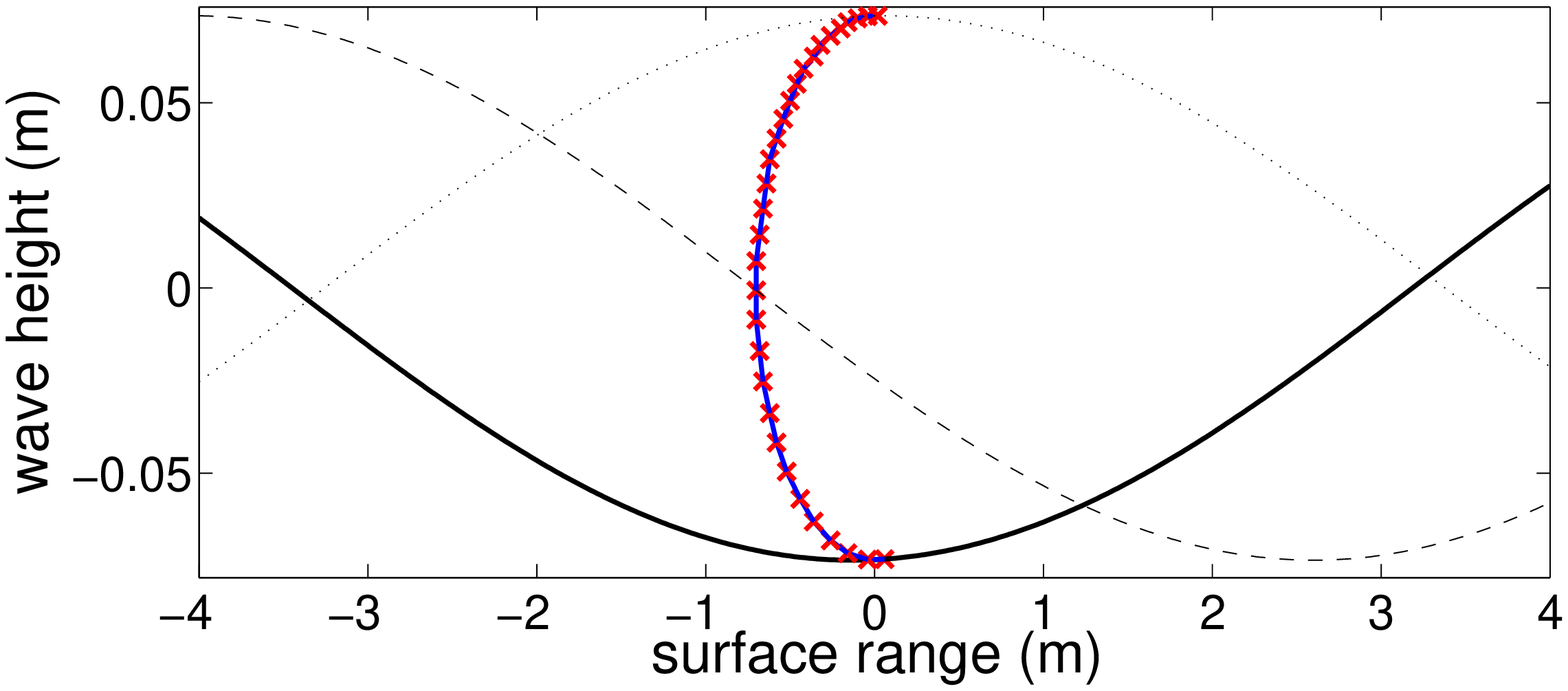,width=9cm}}
  \vspace{0.07cm}
  \centerline{(a)}
\end{minipage}
\begin{minipage}[b]{.99\linewidth}
  \centering
  \centerline{\epsfig{file=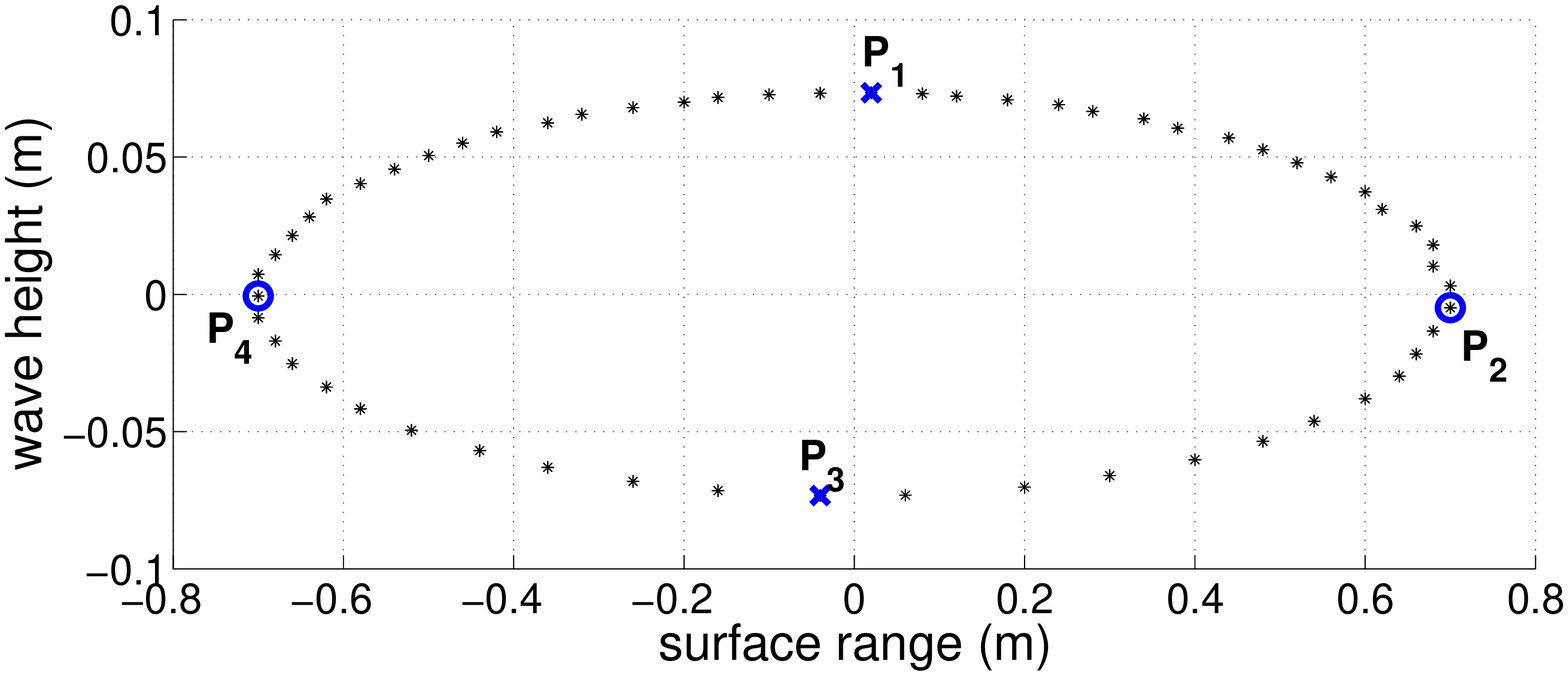,width=9cm}}
  \vspace{0.05cm}
  \centerline{(b)}
\end{minipage}
\caption{Evolution of the spatial positions of the specular points for (a) a half period and (b) a full period. In (a) the
shape of the sinusoid surfaces (dashed and plain black lines) are presented at different times (beginning, middle and end of the
half period) to show the coincidence with the specular point positions (red cross). In (b) the position of the specular points is
given using black dots and particular postions are highlighted using blue circles and blue cross.} 
\label{fig:distortion}
\end{figure}

Specifically, the specular point interpretation offers a more thorough and
intuitive analysis of the sinusoid feature in the time-frequency domain (see
Figure \ref{fig:sinus_3_zoom}). Indeed, the Figures \ref{fig:distortion} show
the different positions of the central specular point for a half and a full
period. The cyclic movement of the specular point clearly appears. It should
nevertheless be noted that the motion of the specular point follows an
elliptical path with a non-uniform speed. The specular point travels at a lower
speed in the vicinity of the top of the sinusoidal sea surface ($P_1$) than near
the bottom ($P_3$).

Finally, the non-linear variations of the velocity when the specular point goes
over the minima or the maxima of the sinusoidal surface is the reason for the
time-frequency feature corresponding to a distorted sinusoid: $t_2-t_1>t_3-t_2$.

\subsection{Surface made of two sinusoids}

Obviously, the monochromatic model of the sea surface is far from
any acceptable level for a realistic interpretation. According to the spectral
model of the sea surface previously presented (see Section
\ref{sec:sea_model}), a sea is characterized by relevant physical phenomena at various scales:
the surface with large wavelength and high amplitude stands for the gravity wave
(large-scale roughness) and the surface with smaller wavelength and amplitude
stands for the capillary and short gravity waves (small-scale roughness).

\begin{figure}[!hbt]
\begin{minipage}[b]{.99\linewidth}
  \centering
  \centerline{\epsfig{file=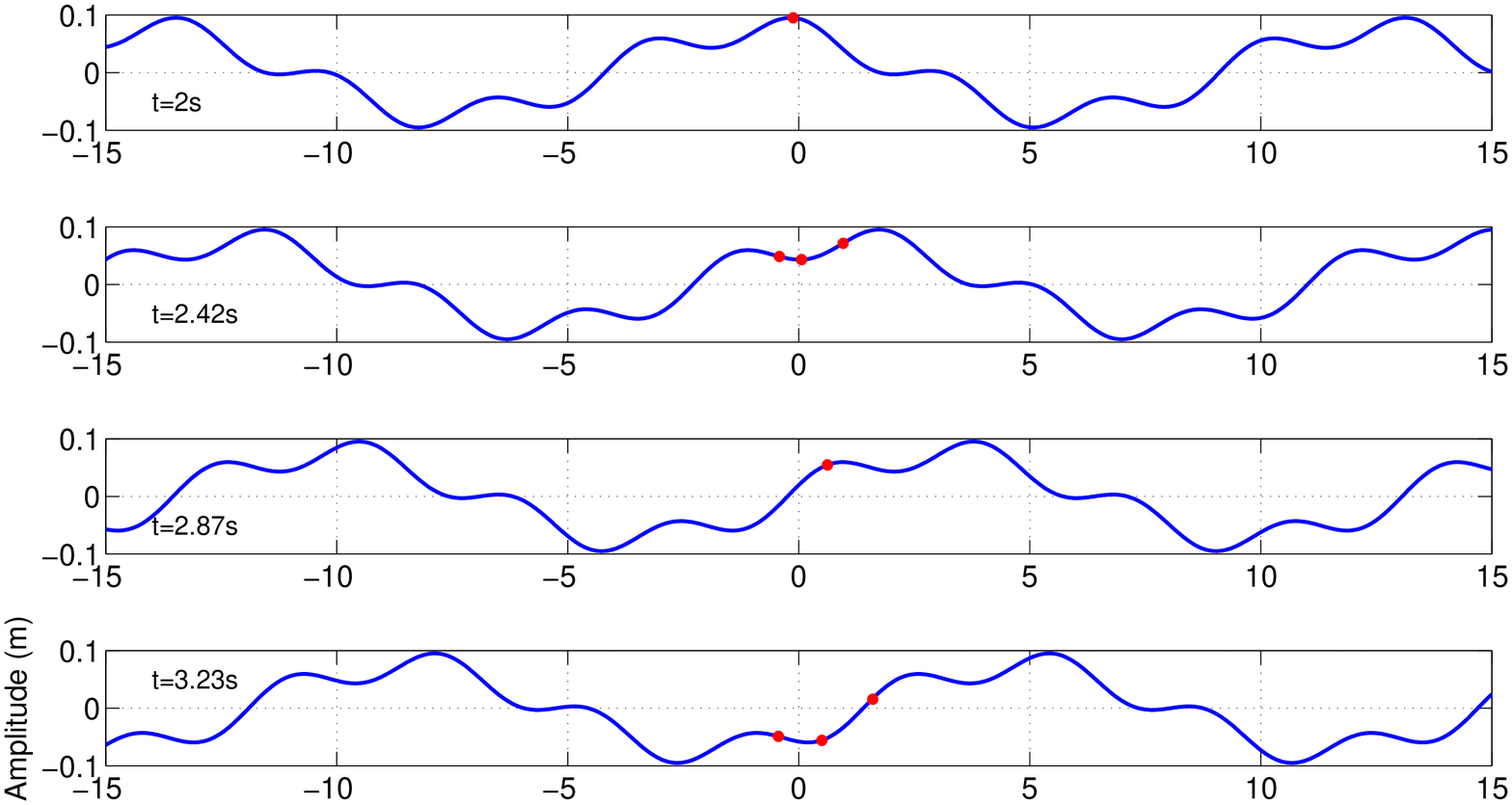,width=9.5cm,trim=1cm 1cm 1cm 1cm, clip=true}}
\end{minipage}
\begin{minipage}[b]{.99\linewidth}
  \centering
  \centerline{\epsfig{file=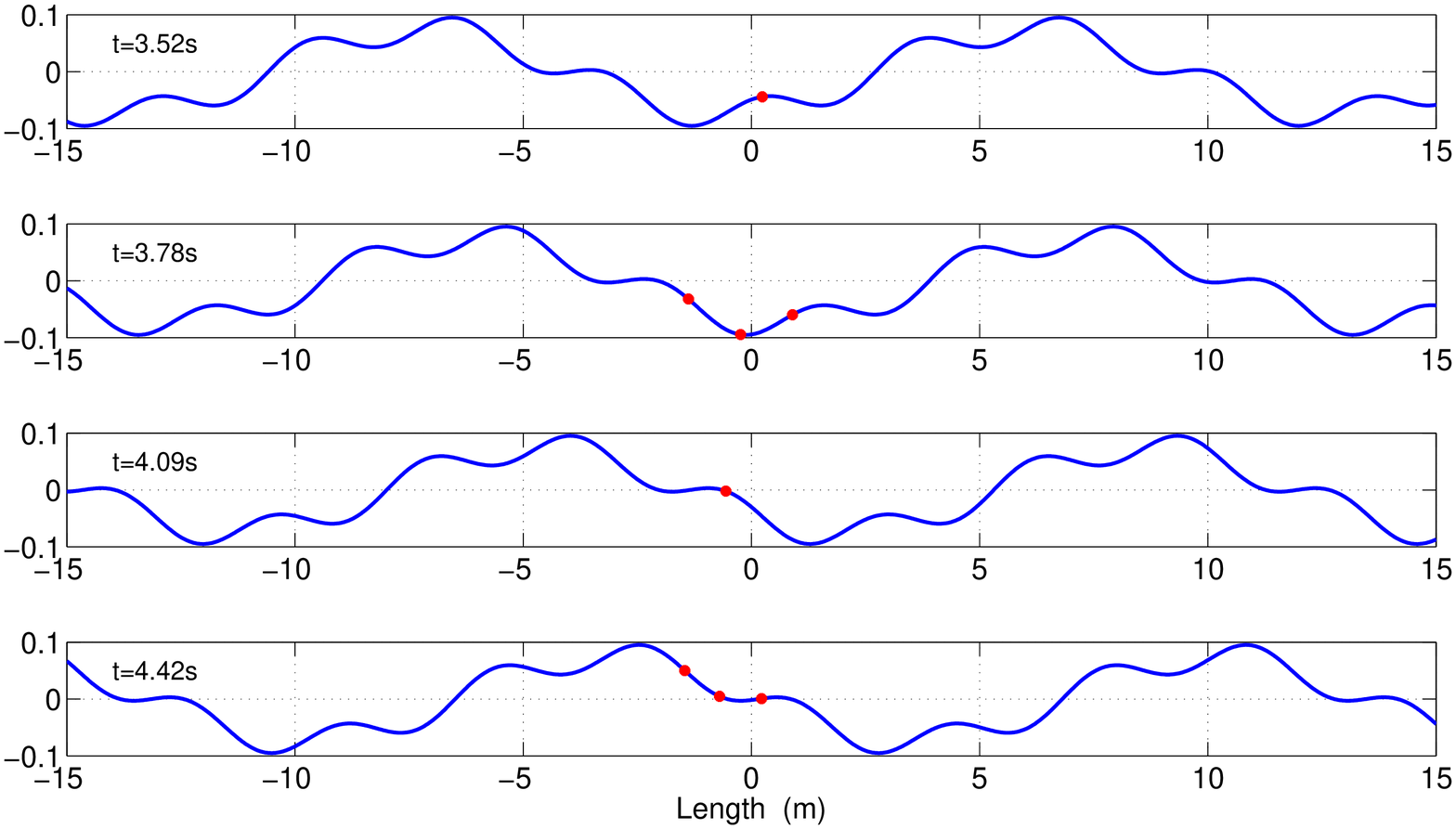,width=9.5cm,trim=1cm 1cm 1cm 1cm, clip=true}}
\end{minipage}
\caption{Sinusoidal surfaces and positions of the specular points at several
times.} 
\label{fig:2sinus_surf}
\end{figure}

In order to come closer to reality, we consider the same previous sinusoidal
surface to which a second sinusoid with smaller amplitude and wavelength is
added. Figure \ref{fig:2sinus_surf} shows examples of such surfaces for a
Beaufort scale 3. For this new generated surface, the parameters of the second
sinusoid correspond to the amplitude of the first divided by three, the
velocity remains the same and the wavelength is divided by four.

\begin{figure}[!ht] 
\centering
\includegraphics[width=\textwidth,trim=1.5cm 1.25cm 1.5cm 1.5cm, clip=true]{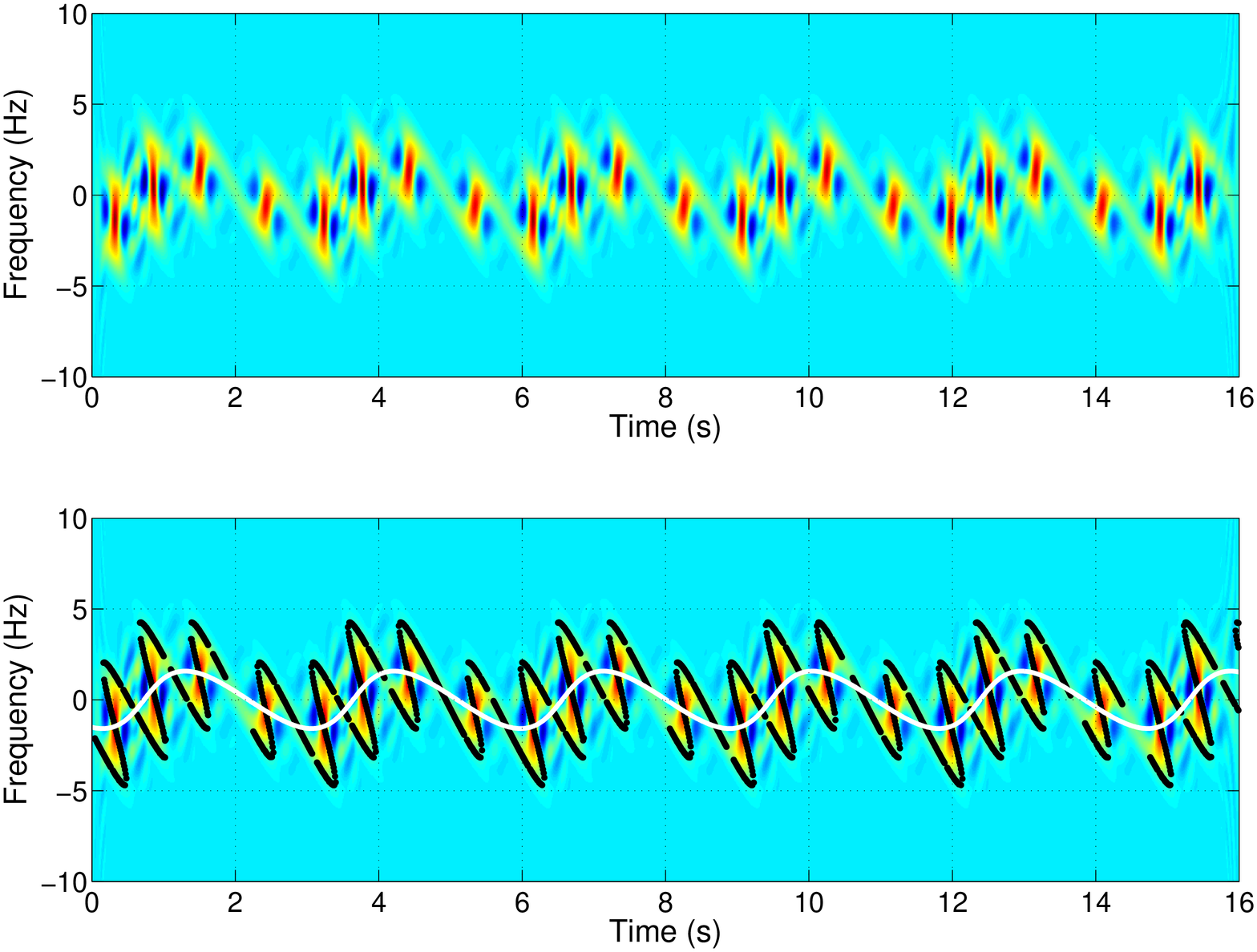}
\caption{(top) TFR (with same color scale) of the signal reflected by the surface composed of two added 
sinusoids and (bottom) previous image including (black line) the Doppler of the
specular points and (white line) the Doppler of the specular point of the
corresponding simple sinusoidal surface (Figure \ref{fig:sinus_spec_3}).}
\label{fig:2sinus_A3}
\end{figure}

\begin{figure}[htb] 
\centering
\includegraphics[width=10cm]{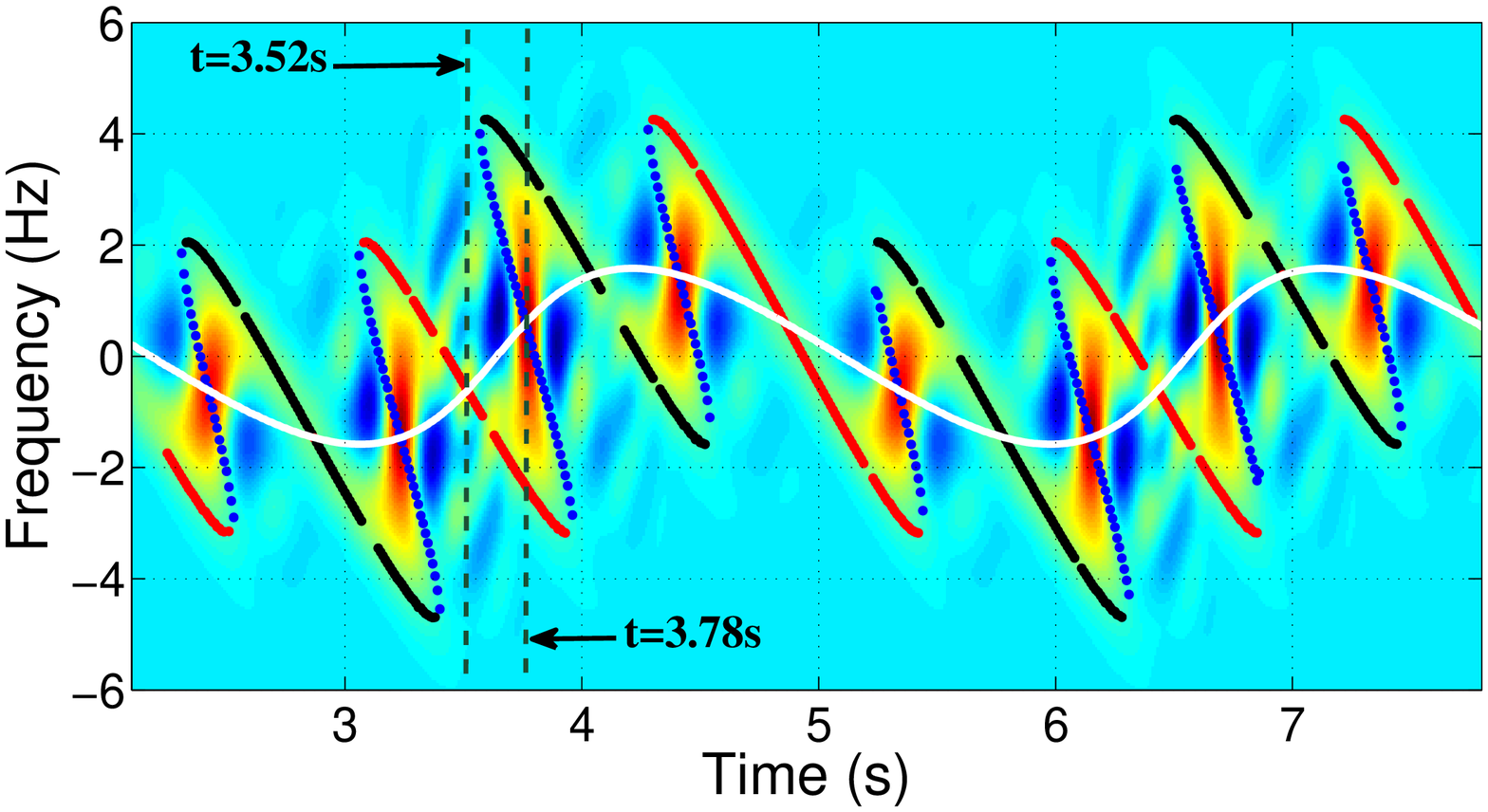}
\caption{Zoom into Figure \ref{fig:2sinus_A3}.b. Highlighted by the black, blue and red dots
is the Doppler associated to each specular point and at each time.}
\label{fig:2sinus_spec_zoom}
\end{figure}

The time-frequency representation of the simulated signal obtained is given in 
Figure \ref{fig:2sinus_A3}. In the same manner as outlined above, the two
sinusoid models can lead to a specular point interpretation. Nevertheless, the
sum of sine curves with different wavelengths involves local curvature
phenomena and several specular points may appear in the central area.
In Figure \ref{fig:2sinus_A3} (bottom) the Doppler
frequencies associated to each specular point are superimposed (in black)
on the time-frequency distribution. It is worth noting that they seem to generate a global continuous Doppler curve in the
time-frequency domain. In Figure \ref{fig:2sinus_spec_zoom} we have highlighted the Doppler frequency associated with the present
specular points at each time with different color. Moreover, the same color is used to highlight the motion of a given specular
point throughout a period.

In Figure \ref{fig:2sinus_A3} (bottom) the evolution of the Doppler frequency due to the specular point in the monochromatic sea
model (same as Figure \ref{fig:sinus_spec_3}) is also added in white. The Doppler curve due to the two sinusoids can be seen as
the so-called
\cite{Chen_06,Thay_04} micro-Doppler phenomenon (compare to the Doppler phenomena
associated with the global signature due to the single sinusoid with larger
amplitude). It appears clearly that there is a global trend highlighted by the
white curve which is due to the motion associated with the main sinusoidal
component (monochromatic model). In addition, the association of the
two sinusoidal components induces oscillatory variations around this main trend.

In the present case, since the ratio between both sinusoids is an integer,
the sea surface considered is deterministic and periodic. So, the TFR is also
periodic. Let us consider a zoom into the TFR focusing on one period (see Figure
\ref{fig:2sinus_spec_zoom}). This can be associated with Figure
\ref{fig:2sinus_surf} which shows the surface at several times and the position
of the corresponding specular points (red star). 

Closer examination of the time-frequency representation reveals the number of specular points and their motion. For instance, at
the time $t=3.78$s (see Figure \ref{fig:2sinus_spec_zoom}), three different intersections exist on the Doppler curve that can be
related to the time-frequency components of three specular points, see Figure \ref{fig:2sinus_surf}. On the other hand, at the
time $t=3.52$s, there is
only one intersection on the Doppler curve, and there exists only one specular
point, see Figure \ref{fig:2sinus_surf}.

\begin{figure}[htb] 
\centering
\includegraphics[width=10cm]{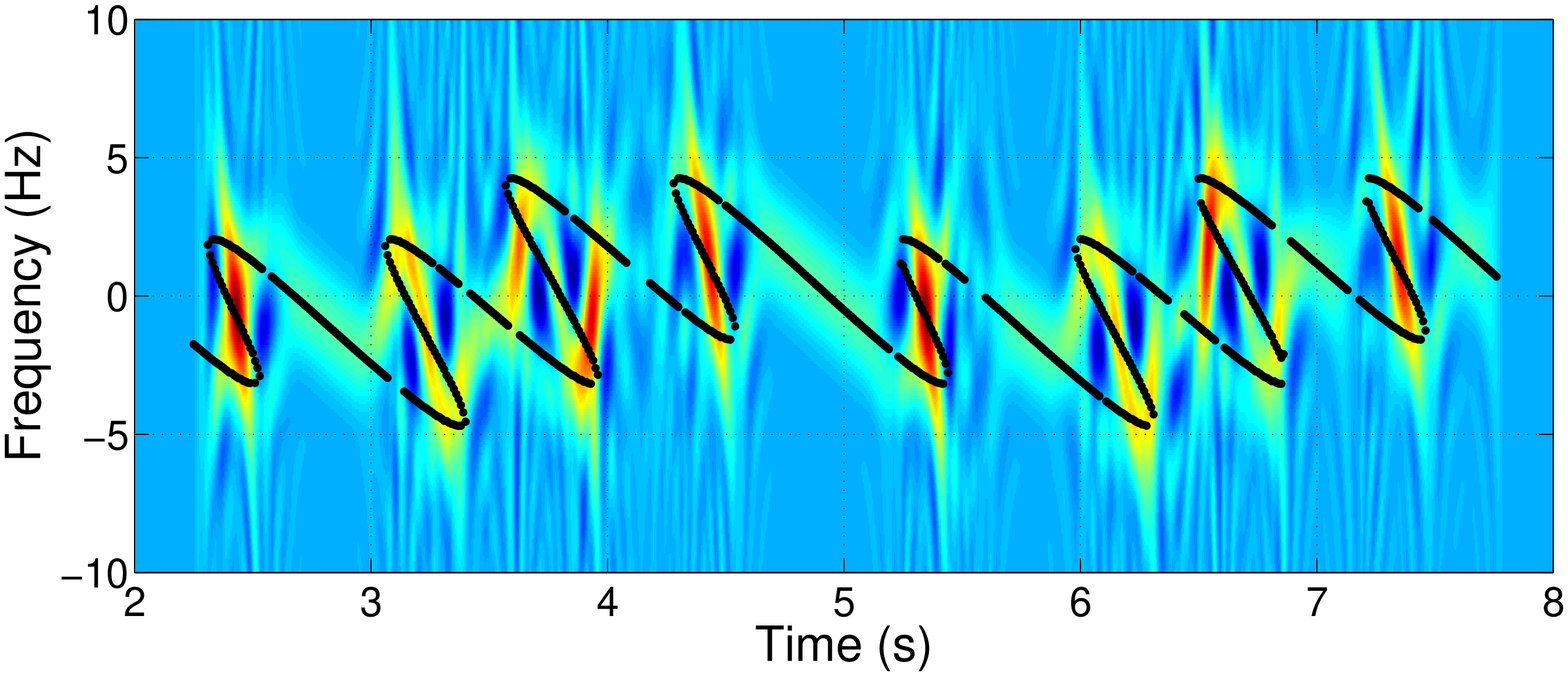}
\caption{Zoom of the time-frequency representation of the received reflected
signal by a surface made of the sum of two sinusoids using the point source
model.}
\label{fig:2sinus_scat}
\end{figure}

Despite the good agreement between MoM and the specular point approximation,
it is important to add that the comparison must not be pushed too far in the
case of a sea surface with multi sinusoid components. In the same way as
for the monochromatic sea surface (see Figure \ref{fig:simu_scat}), Figure
\ref{fig:2sinus_scat} shows the TFR of the electromagnetic field computed using
the specular point source model. Overall, this image has many features in common
with Figure \ref{fig:2sinus_A3}. However, significant differences can be pointed
out between both images. This can be explained largely by the fact that
multi sinusoid components involve local curvatures upon the sea surface. So,
the electromagnetic scattering induces multi path reflection and more complex
interactions that are not taken into account by the specular point
approximation.

\subsection{Surface made of three sinusoids}

With a view to improving our sea surface model, the surface considered now
consists of the sum of three sinusoids with different
wavelengths and amplitudes and above all with different velocities. It is
still considered a surface with parameters corresponding to Beaufort scale 3.
The two added sinusoids are fixed so that the wavelength is respectively divided
by 3 and 5, the amplitude is divided by 2 and 5 and the velocity is multiplied
by 1.25 and 1.6.

In previous cases, the movement of the sea can be summarized into a
global translation at a constant speed, see Figures \ref{fig:sinus_surf_3} and
\ref{fig:2sinus_surf}. In reality, the sea surface has to be considered as a
dispersive medium for the sea waves (speed depends upon the wavelength, see
equation (\ref{e:dispersion})). This dispersion induces deformations over
time in addition to the translation. This last issue introduces new
physical phenomena and modifies the TF features. 

\begin{figure}[htb] 
\centering
\includegraphics[width=\textwidth,trim=1.25cm 0.75cm 1.25cm 1.25cm, clip=true]{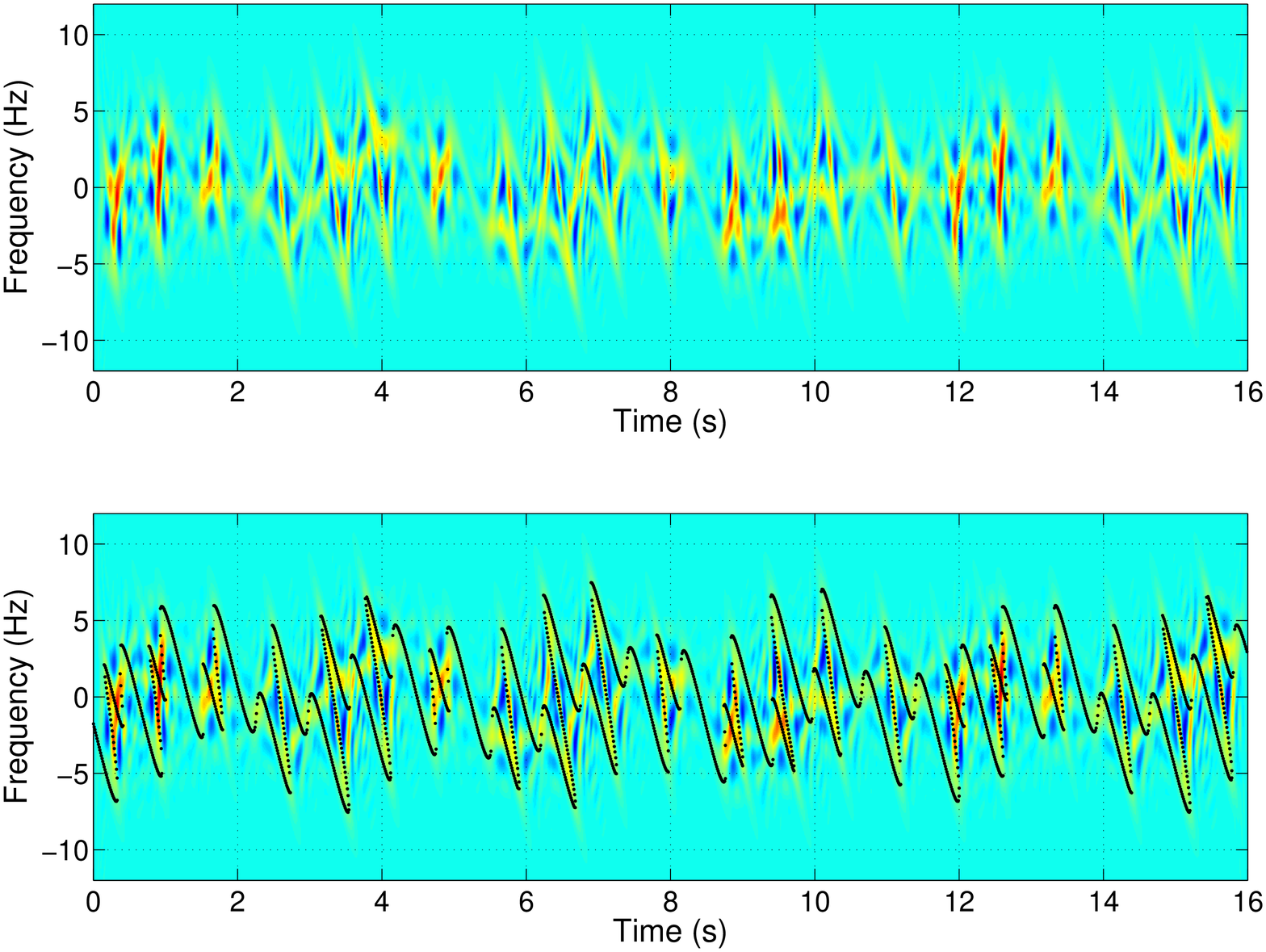}
\caption{TF representation of the signal reflected from a surface made of the
sum of three sinusoids with different velocities. In the bottom image the
Doppler frequency due to the specular points is added (black dots).}
\label{fig:TF_3sinus}
\end{figure}

The TFR obtained for the three sinusoid models is presented in Figure
\ref{fig:TF_3sinus}. As in the case of two sinusoid models, the TFR allows us
to view the micro-Doppler phenomena (oscillatory variations around the main
trend). Nevertheless, the structure of these oscillations seems slightly more
complex, in the present case.

\begin{figure}[htb] 
\centering
\includegraphics[width=\textwidth,trim=1.25cm 1.25cm 1.25cm 1.25cm, clip=true]{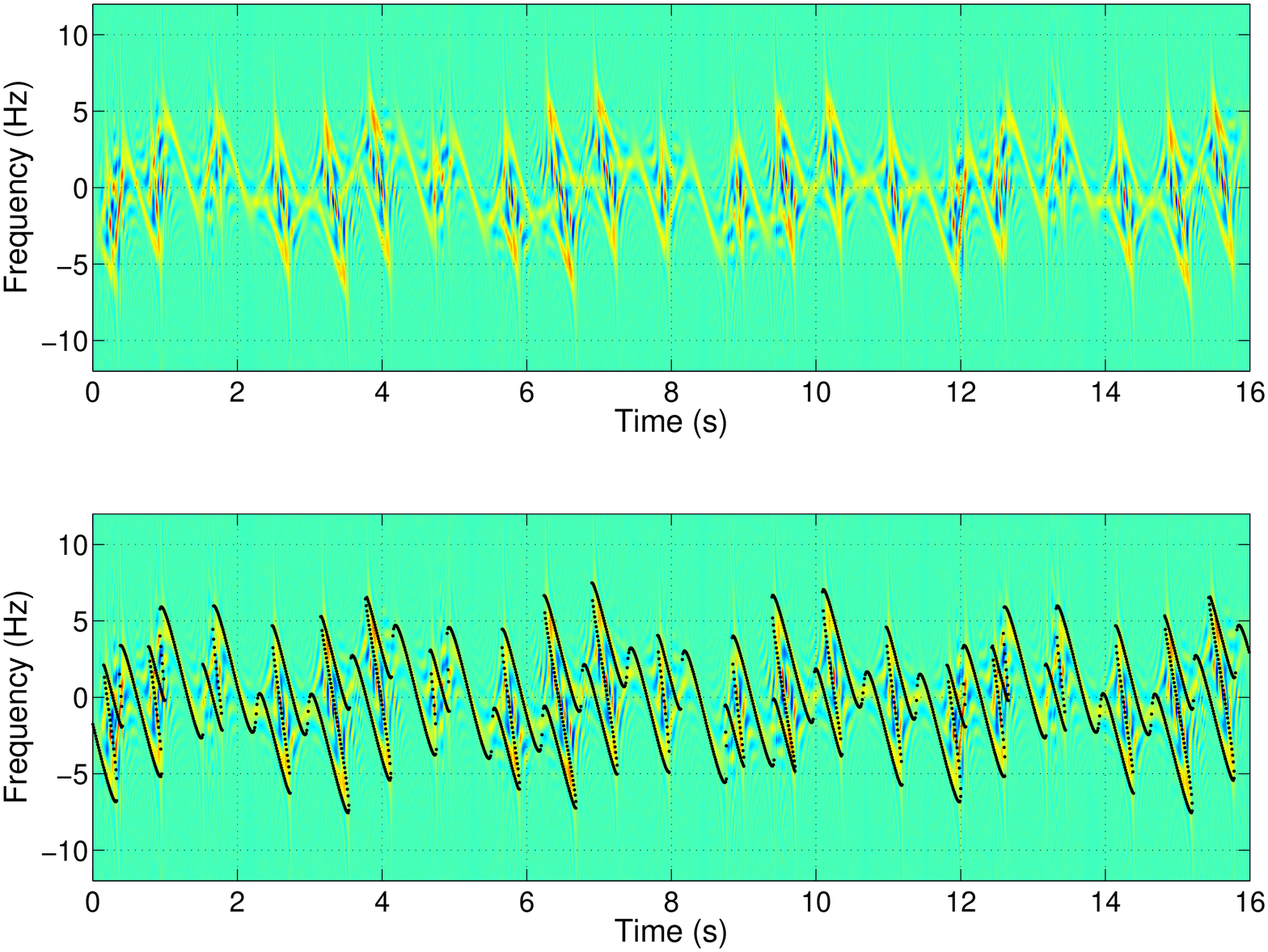}
\caption{TF representation of the signal (point source approach) reflected from
a surface made of the sum of three sinusoid with different velocities. In the
bottom image the Doppler frequency due to the specular points added using black
dots.}
\label{fig:TF_3sinus_scat}
\end{figure}

Once again, to understand the TFR, we can make use of the specular point
approximation. Figure \ref{fig:TF_3sinus_scat} shows the TFR of the
electromagnetic field approximated with the specular point source model.
Despite significant differences, we do see that the global shape is very
similar to Figure \ref{fig:TF_3sinus}. Therefore the
specular point approximation can, to large extent, provide a relevant
description.

\begin{figure}[!hbt]
\begin{minipage}[b]{.99\linewidth}
  \centering
  \centerline{\epsfig{file=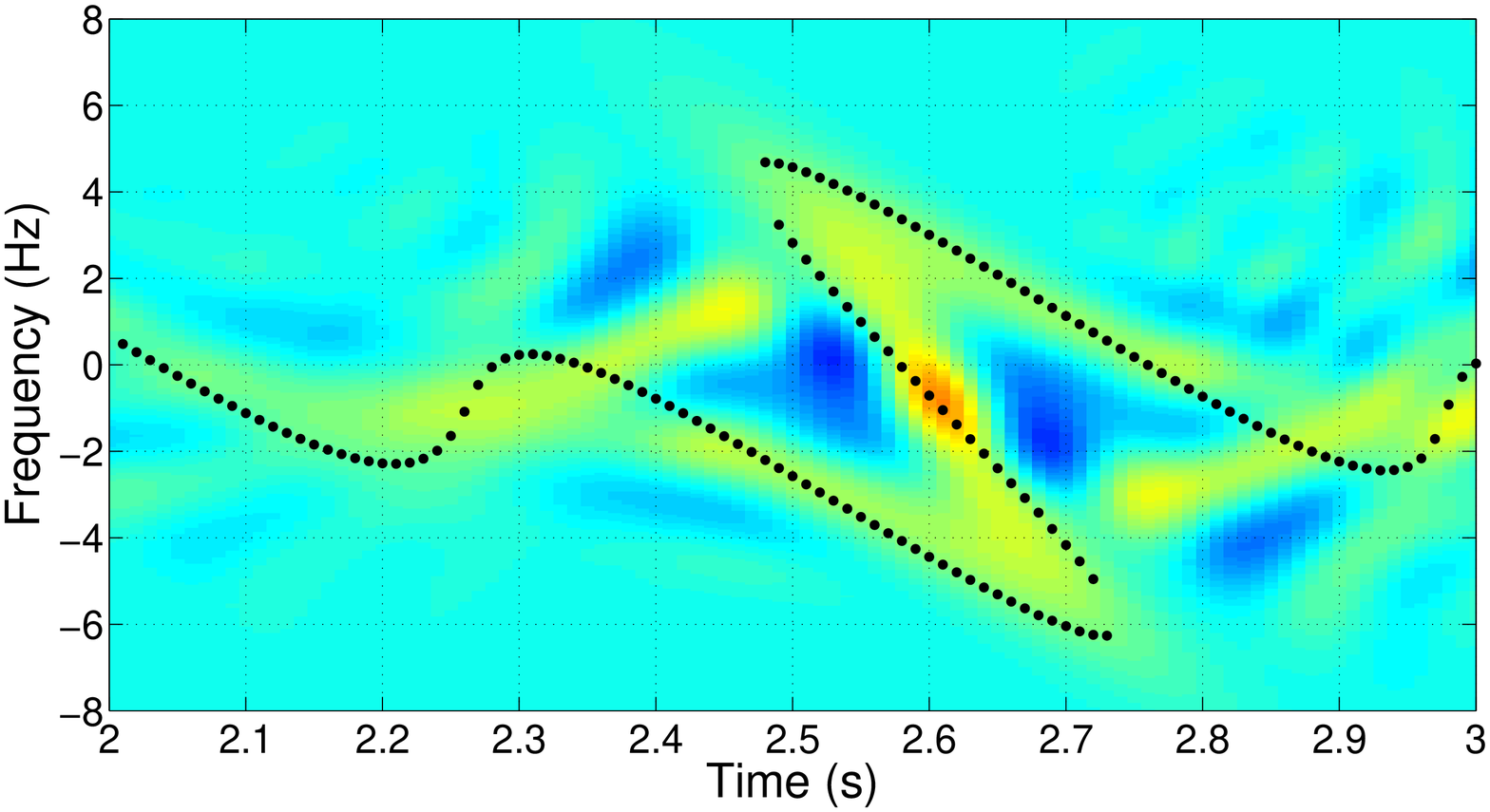,width=8cm}}
\end{minipage}
\begin{minipage}[b]{.99\linewidth}
  \centering
  \centerline{\epsfig{file=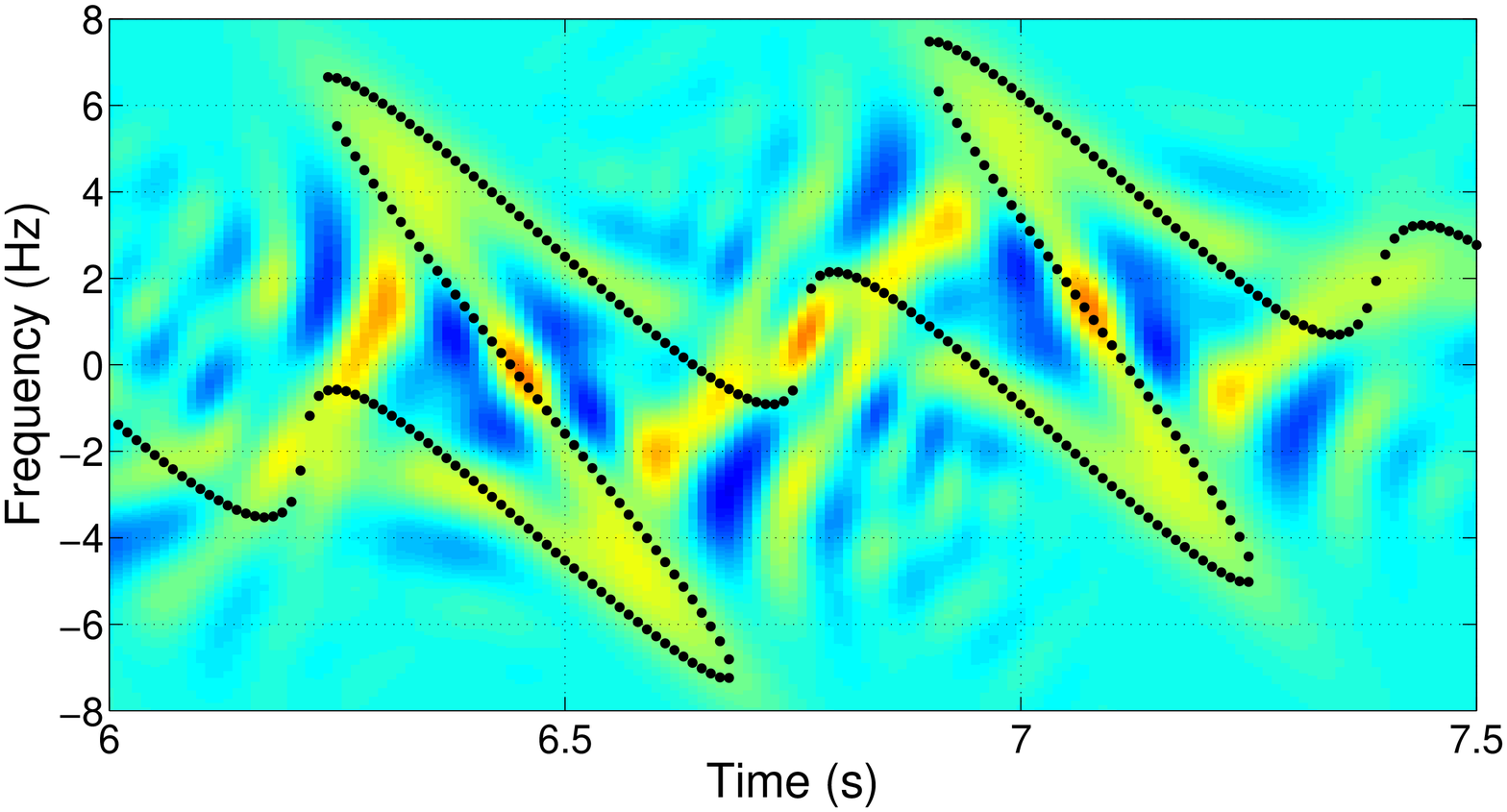,width=8cm}}
\end{minipage}
\caption{Zoom on an area of interest in Figure \ref{fig:TF_3sinus}.} 
\label{fig:TF_3sinus_zoom}
\end{figure}

\begin{figure}[htb] 
\centering
\includegraphics[width=9cm,trim=1.25cm 1.25cm 1.25cm 1.25cm, clip=true]{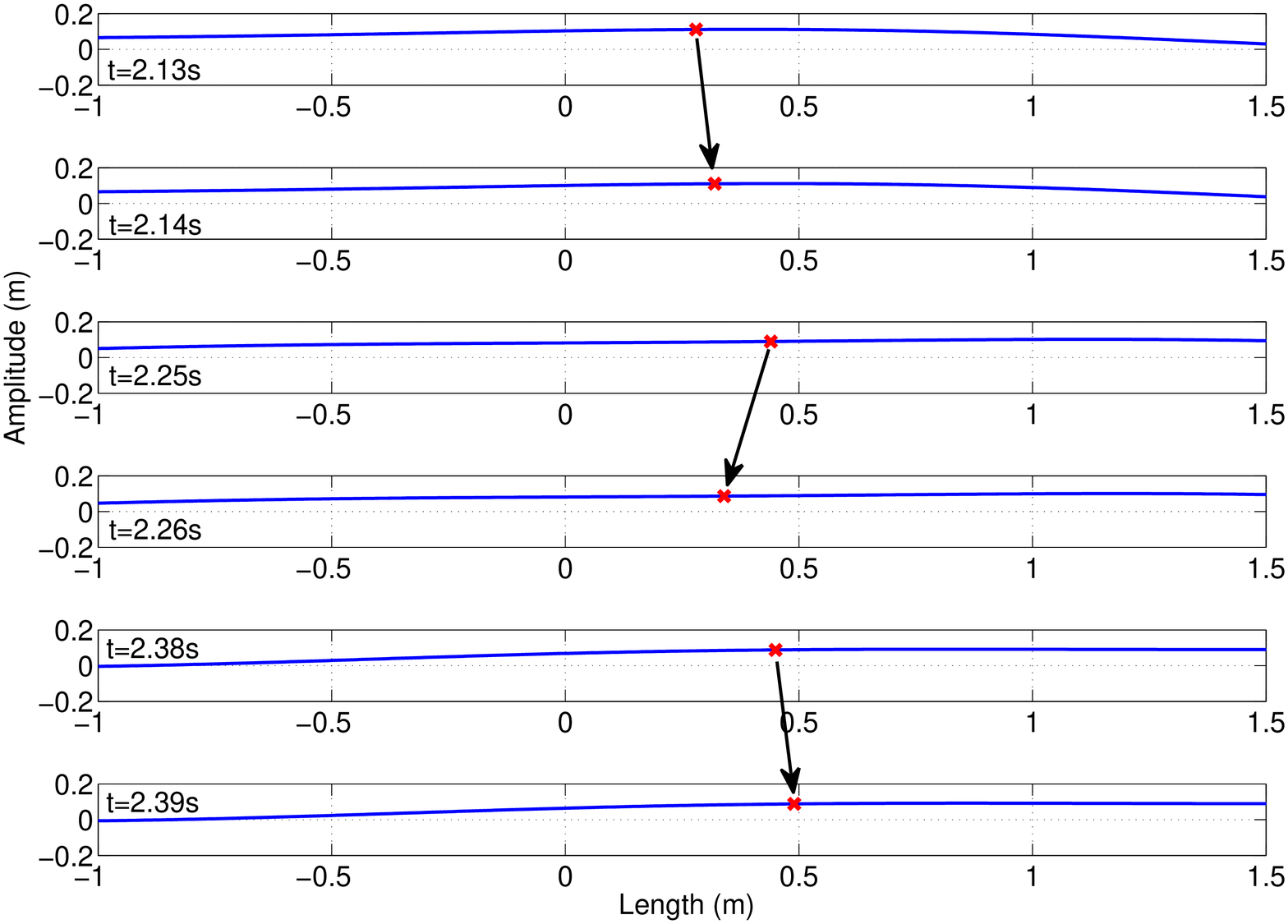}
\caption{Three sets of  two (zoom) successive surfaces (blue line) and the evolving positions (shown by the black arrows) of the
specular point (red star).}
\label{fig:TF_3sinus_surf1}
\end{figure}

\begin{figure}[htb] 
\centering
\includegraphics[width=9.25cm,trim=1.5cm 1.35cm 1.25cm 1.25cm,clip=true]{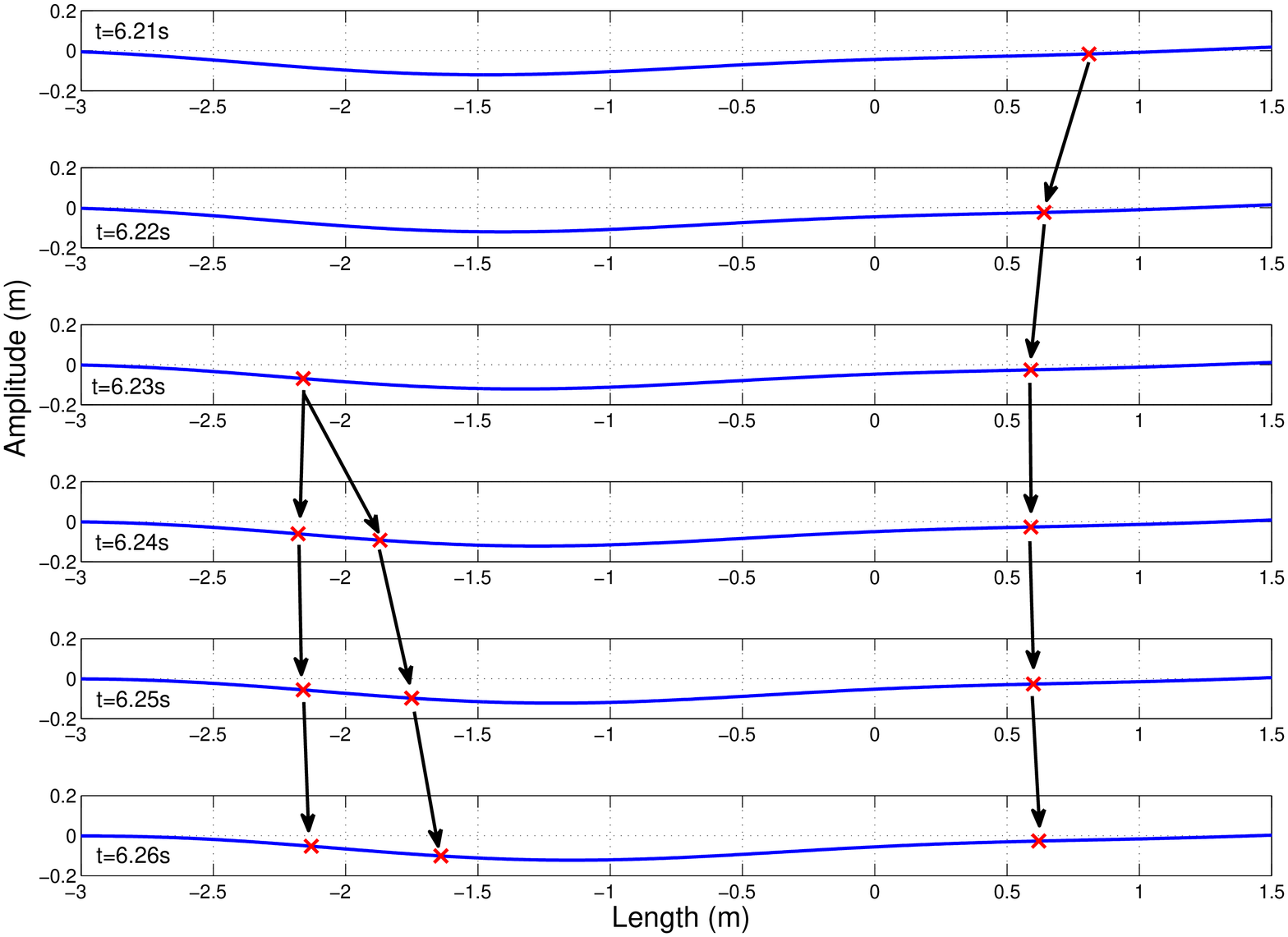}
\caption{Example showing (zoom) six successive surface shapes (blue line) and the evolving positions (shown by the black arrow) of
the specular points (red star).}
\label{fig:TF_3sinus_surf2}
\end{figure}

The TF representation for three sinusoid sea surfaces brings about quite
complex phenomena that deserve more sustained analysis. Thus 
Figure \ref{fig:TF_3sinus_zoom} shows zooms of two areas of interest in Figure
\ref{fig:TF_3sinus}: from $t=2$s to $t=3$s and from $t=6$s to $t=7.5$s.
In addition, Figures \ref{fig:TF_3sinus_surf1} and \ref{fig:TF_3sinus_surf2}
show the sea surface with the specular points at several successive times in
these two periods.

Between $t=2.13$s and $t=2.39$s there is only one specular point
inducing a Doppler curve that appears to be oscillating. In Figure
\ref{fig:TF_3sinus_surf1}, we can see that in a first step
($t=2.13$s/2.14s) the specular point moves away from the receiver with
a speed greater than the global sea movement. Then Doppler frequency is
negative and lower than that obtained for a harmonic sea surface.
In a second step ($t=2.25$s/2.26s) the specular point locally tends to
move closer to the receiver. The Doppler frequency increases and even reaches a
positive maximum. In the last step ($t=2.38$s/2.39s), the specular point
speeds up the shift to the right, and Doppler frequency returns to negative
domain. This oscillation cannot be explained by a global translation of the sea
surface but must be seen as the consequence of the local sea surface
deformations related to dispersion.

The deformations due to dispersion increase the complexity of the local
curvature observed at the sea surface. Between $t=6.21$s and $t=6.26$s, the
number of specular points changes over time. Figure
\ref{fig:TF_3sinus_surf2} shows that there is only one specular point from
$t=6.21$s to $t=6.22$s. Then, a new specular point appears at $t=6.23$s.
Finally this new point forks into two new specular points. Both specular points move closer to the observer but with
different velocities. In the same way, we can see that dispersion and the local deformations of the
sea surface periodically lead to the disappearance of one or more specular
points. Somehow we must highlight the fact that the Doppler curves associated
to each specular points still form a continuous Global Doppler curve.

\section{Reflected signal in the TF domain - realistic sea surfaces}\label{sec:sea}

After considering the scattering by canonical surfaces, in this section, we investigate the TFR of the electromagnetic signal
obtained for the
scattering by realistic sea surfaces. Clearly, the surfaces generated using the
theory presented in Section \ref{sec:sea_model} are far more complex that those
obtained using sinusoids or the sum of several sinusoidal functions.
However, we are going to find most of the
previous expounded phenomena that influence the time-frequency features.

\subsection{TFR of signals reflected by several sea surfaces}

\begin{figure}[!ht] 
\centering
\includegraphics[width=\textwidth]{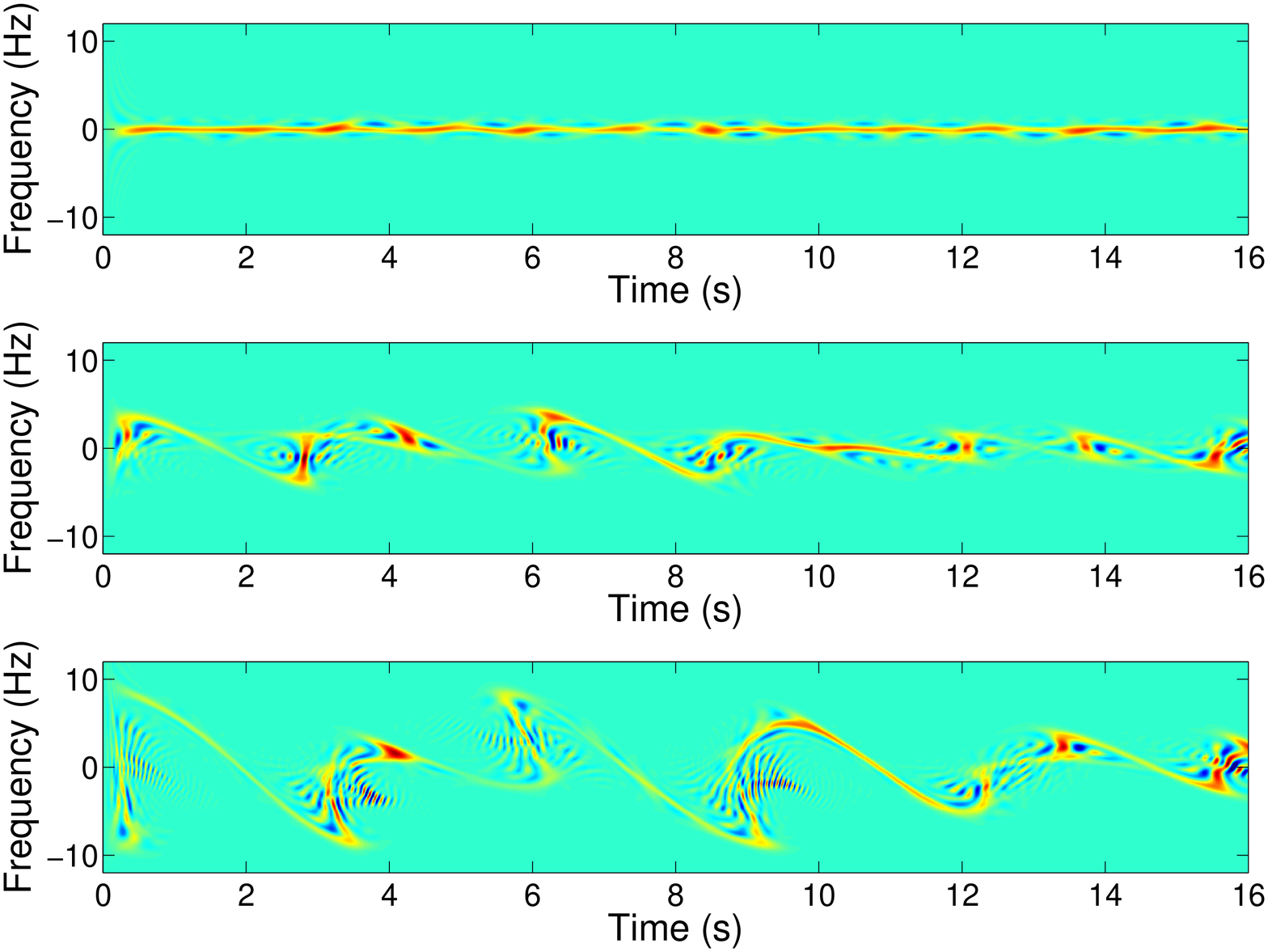}
\caption{TF representation of the signal reflected from a sea state with scale (top to buttom) 2, 3 and 4.}
\label{fig:TF_mer_234}
\end{figure}

Figures \ref{fig:TF_mer_234} shows the TFR obtained from the signal reflected
from sea surfaces with respectively Beaufort scale 2, 3 and 4. For the sea
surface with Beaufort scale 2, from the receiver point of view, the surface is
almost flat which is why the time frequency feature is mainly focused
upon the zero-Doppler frequency line.
The oscillations around this strip, which can be seen as micro-Doppler phenomena,
are linked with the small oscillation of the surface. For the two other sea
states, the surface can no longer be considered to be almost flat and then the
TF feature appears as a far more structured geometry.

\begin{figure}[htb] 
\centering
\includegraphics[width=\textwidth]{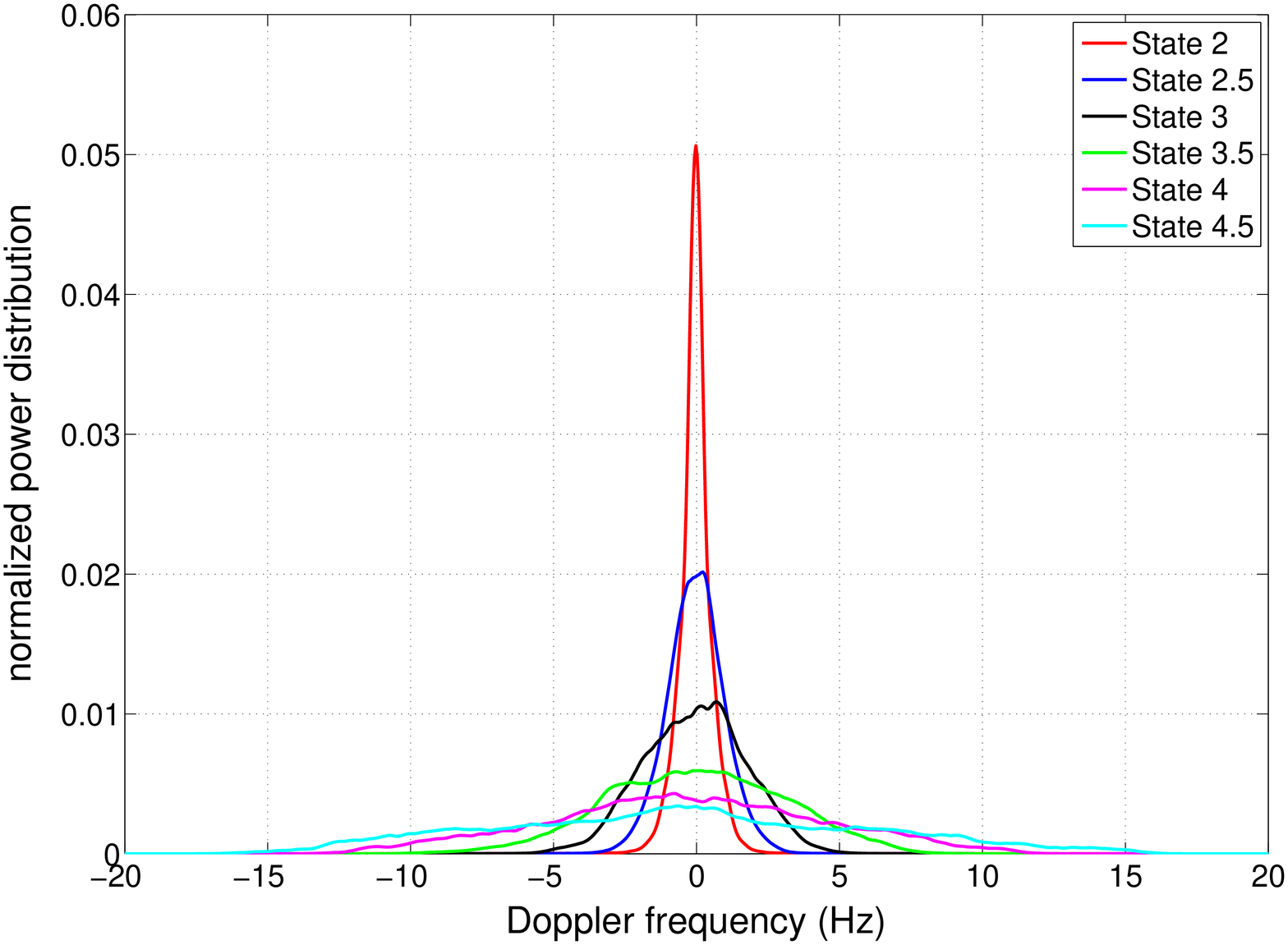}
\caption{Normalized power distribution for several sea states.}
\label{fig:dist_nrj}
\end{figure}

These TF representations suggest that there is great potential for
feature extraction and specialized remote sensing applications.
At first glance, we can see that the spreading of the Doppler frequency is in
close conjunction with the sea state. This is illustrated in figure \ref{fig:dist_nrj} which shows the (normalized) power distribution according to the Doppler frequency obtained from several sea states. It can also be notated that the micro-Doppler signature is directly related to the surface roughness and the fluid dynamics of the sea surface.

\subsection{Analysis of the TFR for the sea surface at scale 3}
\begin{figure}[!ht] 
\centering
\includegraphics[width=\textwidth,trim=1.25cm 1.25cm 1.25cm 1.25cm, clip=true]{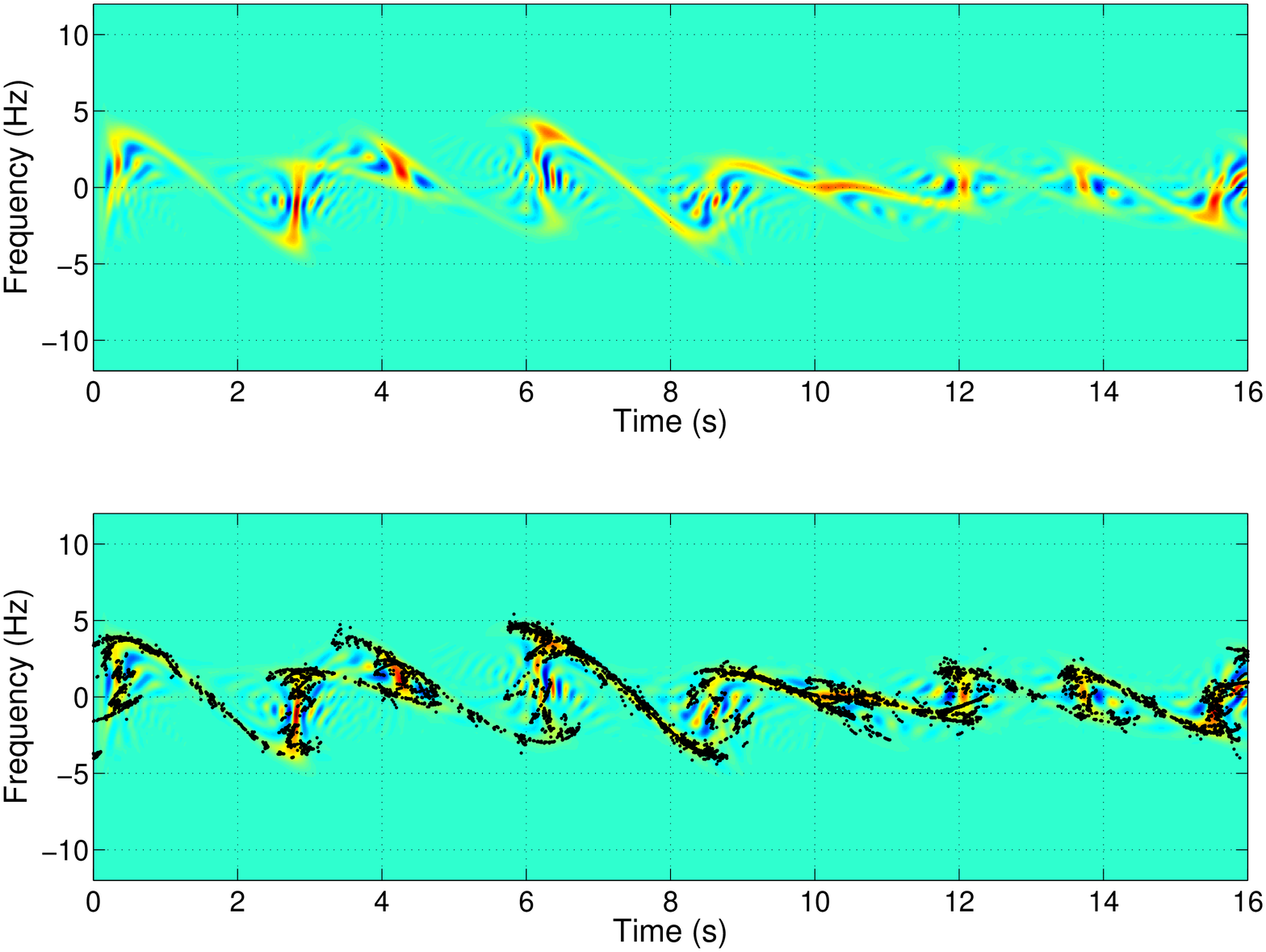}
\caption{TF representation of the signal reflected from a sea surface with state scale 3 and the frequency Doppler of the
specular points (black dots).}
\label{fig:TF_mer_3_spec}
\end{figure}

Figure \ref{fig:TF_mer_3_spec} shows the TFR and (stacked) the evolution (blue
points) of the Doppler frequencies of the specular points. The Beaufort scale of
the sea surface considered is 3. 

It is clear that the understanding of this TFR image is much more complicated
than canonical cases. Moreover, it is difficult to assert whether the
Doppler curves related to each specular point form a continuous global curve in a present case.
However, the cloud of points corresponding to each Doppler specular point bears
similarities with the Doppler curve obtained in the two or three sinusoid
surface cases.

\begin{figure}[!htb] 
\centering
\includegraphics[width=12cm]{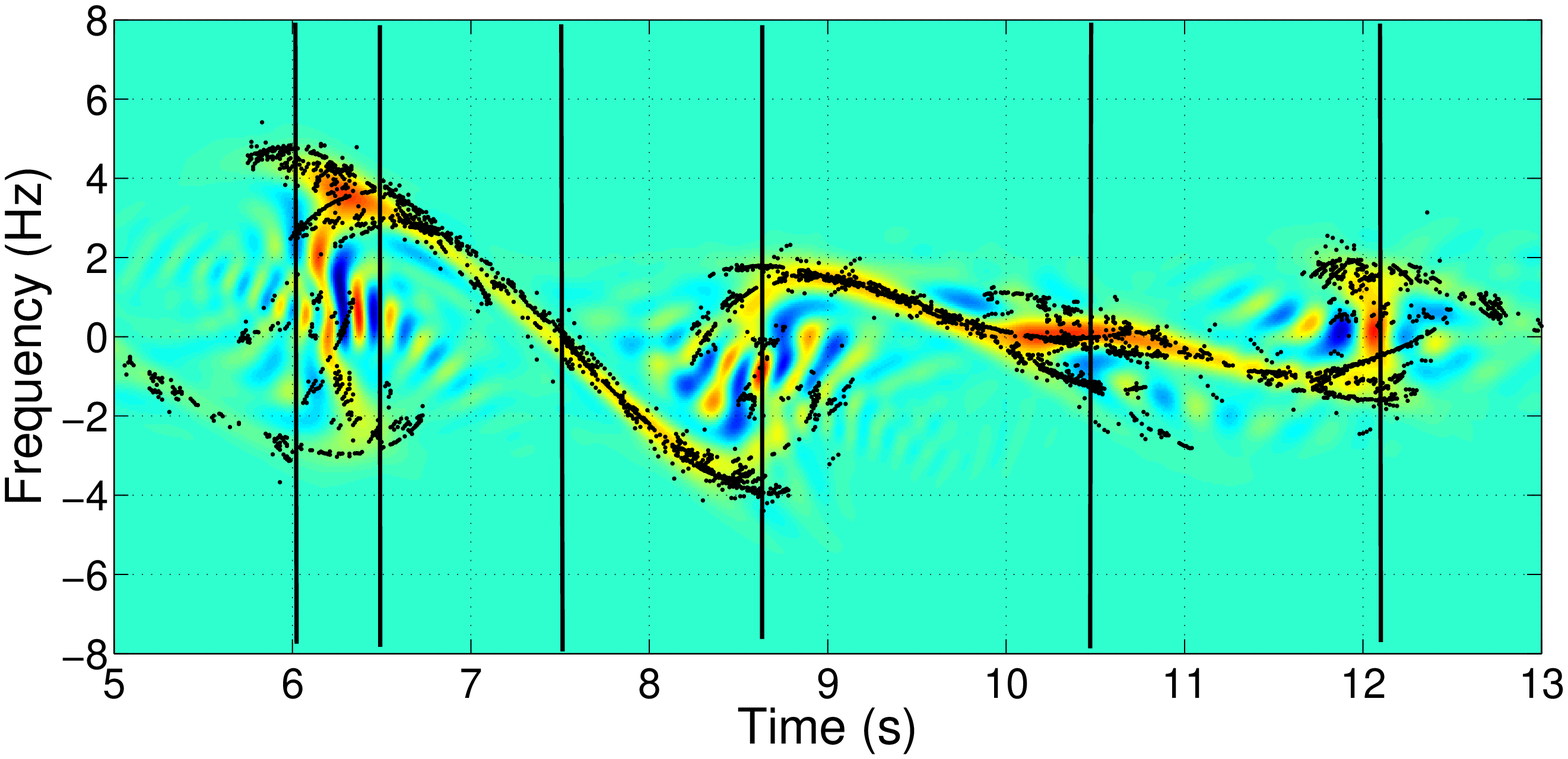}
\caption{Zoom of Figure \ref{fig:TF_mer_3_spec}. Black lines show the times
corresponding to the sea surface drawn in Figure \ref{fig:TF_mer_3_surf_spec}.}
\label{fig:TF_mer_3_spec_b}
\end{figure}

\begin{figure}[!htb] 
\centering
\includegraphics[width=9.2cm,trim=1.5cm 1.5cm 1.5cm 1.5cm, clip=true]{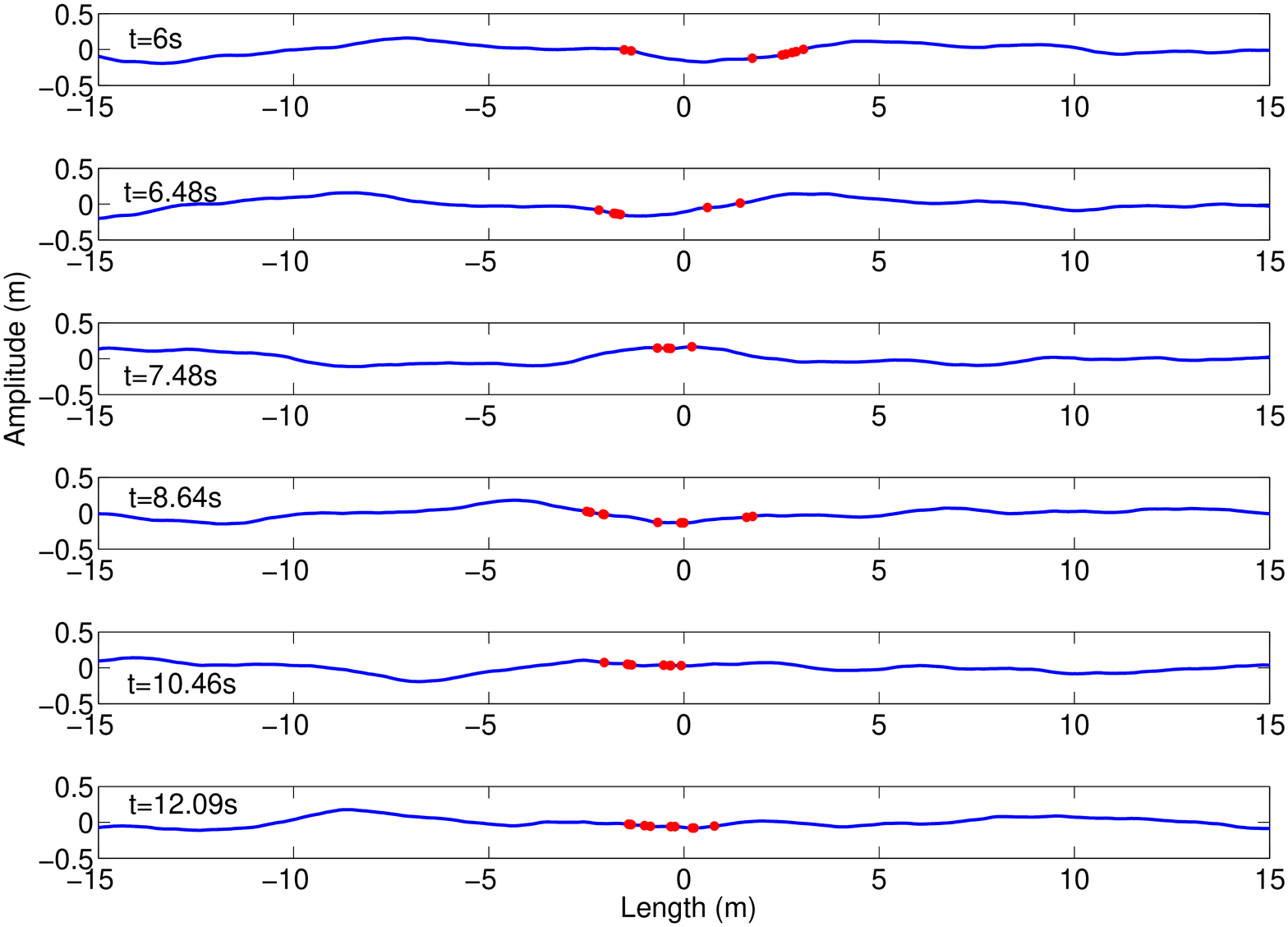}
\caption{Sea surface at several times and in red the positions of the specular points.}
\label{fig:TF_mer_3_surf_spec}
\end{figure}

In order to improve readability, Figure \ref{fig:TF_mer_3_spec_b} focuses on a
restricted time domain and Figure \ref{fig:TF_mer_3_surf_spec} represents the
snapshots of the sea surface and the computed specular points (red stars) at specific
times: $t=6$s, $t=6.48$s, $t=7.48$s, $t=8.64$s, $t=10.46$s and
$t=12.09$s.

First of all, it is noticeable that the TFR induced by a realistic sea surface
looks like the Doppler curves observed in canonical cases at certain times
($t=7.48$s for instance). These situations most often occur when specular
points are fewer and form a gathering of points as in Figure
\ref{fig:TF_mer_3_surf_spec}. In these situations, it becomes easy to make
comparisons with the canonical cases previously studied and it may give rise to
the very simple physical interpretations given in Section \ref{sec:sinus}.

Conversely, when the distribution of the specular points spreads over a larger
region (between $t=6$s and $t=6.48$s), the time frequency signature becomes
more complex and is probably the consequence of multi-scale phenomena.  
Furthermore, Figure \ref{fig:TF_mer_3_surf_spec} shows that realistic sea
surfaces often cause non uniform specular point distributions. In certain
cases, the distribution of specular points is structured as different clusters.
In the latter case, the TFR is characterized by multi component electromagnetic
interactions.

In any case, it is reasonable to assume that deterministic or statistical analysis
of the TFR related to a given sea surface could provide significant information
about the fluid dynamics of the sea. In a first instance, the analysis could
consist in an extraction of the global TFR structure (pseudo periodic
structure). In a second phase, the analysis could focus upon the specific TFR
structures related to the multi-scale phenomena. Progress will be made in future works on this matter.

\clearpage

\section{Conclusion}\label{sec:conclusion}
This paper investigates the feature in the time-frequency domain of
L-band signals reflected from sea surfaces. The data used have been generated
using a MoM method and the TFR have been obtained from the TFTB toolbox. It has
been shown that the signatures of the reflected signals in the TF domain are
linked with the physics of the fluid dynamics. These proposed results provide the basis for interpreting the scattering of L-band
waves by time varying sea surfaces. More precisely, it has been shown that the specific patterns (looking like sigmoid
functions) can be linked to the specular reflections of the incident wave on the sea surface. In addition, we have seen
that local oscillations of the time frequency curve could be related to dispersion phenomena of the fluid surface. This
preliminary study opens up new opportunities for GNSS signals (L-band signals) as remote
sensing systems for dynamic sea surfaces.

Future works will focus on the estimation of oceanographic parameters from
these time-frequency analyses of reflected signals. Eventually, this study will later be extended to the analysis of non-linear
sea or breaking waves \cite{Khairi_art_2013}.

\section*{Acknowledgment}
This work was supported by `GIS Europ\^ole Mer' and ENSTA Bretagne.


\end{document}